\documentclass{elsart}

\usepackage{graphics}
\usepackage{graphicx}
\usepackage{epsfig}

\usepackage{amssymb}

\usepackage{url}

\usepackage[usenames]{color}

\begin{document}

\begin{frontmatter}

\title{The Nylon Scintillator Containment Vessels for the Borexino Solar Neutrino Experiment}

\author[pu-chemeng]{J. Benziger},
\author[pu-phys]{L. Cadonati\thanksref{laura}},
\thanks[laura]{Now at Massachusetts Institute of Technology, Cambridge, MA, USA}
\author[pu-phys]{F. Calaprice\corauthref{cor}},
\corauth[cor]{Corresponding author. Tel: +1-609-258-4375; fax: +1-609-258-2496.}
\ead{calaprice@princeton.edu}
\author[pu-phys]{E. de Haas},
\author[pu-phys]{R. Fernholz\thanksref{fernholz}},
\thanks[fernholz]{Now at Kingsley, MI, USA}
\author[pu-phys]{R. Ford\thanksref{richard}},
\thanks[richard]{Now at SNOLab, Sudbury, ON, Canada}
\author[pu-phys]{C. Galbiati},
\author[pu-phys]{A. Goretti},
\author[pu-phys]{E. Harding\thanksref{beth}},
\thanks[beth]{Now at Lockheed Martin Corporation, Sunnyvale, CA, USA}
\author[pu-phys]{An. Ianni},
\author[pu-phys]{S. Kidner},
\author[pu-phys]{M. Leung},
\author[pu-phys]{F. Loeser},
\author[pu-phys]{K. McCarty},
\author[pu-phys]{A. Nelson},
\author[pppl]{R. Parsells},
\author[pu-phys]{A. Pocar\thanksref{andrea}},
\thanks[andrea]{Now at Stanford University, Stanford, CA, USA}
\author[pu-phys]{T. Shutt\thanksref{tom}},
\thanks[tom]{Now at Case Western Reserve University, Cleveland, OH, USA}
\author[pu-phys]{A. Sonnenschein\thanksref{andrew}},
\thanks[andrew]{Now at University of Chicago, Chicago, IL, USA}
\author[pu-phys]{R. B. Vogelaar\thanksref{bruce}}
\thanks[bruce]{Now at Virginia Polytechnic Institute, VA, USA}

\address[pu-chemeng]{Chemical Engineering Department, Princeton University, Princeton, NJ 08544, USA}
\address[pu-phys]{Physics Department, Princeton University, Princeton, NJ 08544, USA}
\address[pppl]{Princeton Plasma Physics Laboratory, Princeton, NJ 08543, USA}

\setcounter{footnote}{0}

\begin{abstract}
Borexino is a solar neutrino experiment designed to observe the 0.86\,MeV
$^7$Be neutrinos emitted in the pp cycle of the sun.  Neutrinos will be
detected by their elastic scattering on electrons in 100~tons of liquid
scintillator.  The neutrino event rate in the scintillator is expected to be
low ($\sim$0.35 events per day per ton), and the signals will be at energies
below 1.5\,MeV, where background from natural radioactivity is prominent.
Scintillation light produced by the recoil electrons is observed by an array of
2240 photomultiplier tubes.  Because of the intrinsic radioactive contaminants
in these PMTs, the liquid scintillator is shielded from them by a thick barrier
of buffer fluid.   A spherical vessel made of thin nylon film contains the
scintillator, separating it from the surrounding buffer.  The buffer region
itself is divided into two concentric shells by a second nylon vessel in order
to prevent inward diffusion of radon atoms.  The radioactive background
requirements for Borexino are challenging to meet, especially for the
scintillator and these nylon vessels.  Besides meeting requirements for low
radioactivity, the nylon vessels must also satisfy requirements for mechanical,
optical, and chemical properties.  The present paper describes the research and
development, construction, and installation of the nylon vessels for the
Borexino experiment. 

\end{abstract}

\end{frontmatter}

\section{Introduction and Overview of the Borexino Experiment}
\label{s:intro}

Borexino \cite{borexinopaper} is a liquid scintillation detector designed to observe solar neutrinos. 
It is located 1.4\,km (3500 meters water equivalent) underground, in the Gran
Sasso National Laboratory in Italy. 
The goals of the project are two-fold: 
to confirm that the sun produces energy in accord with the Standard Solar
Model~\cite{std-solar-model} through the series of
neutrino-yielding nuclear fusion reactions known as the $pp$ cycle, and to 
use the sun as a powerful source of neutrinos to study the phenomena of 
neutrino oscillations~\cite{nuosc1,nuosc2}.  The main observational target
of the detector is the monoenergetic ($E = 862$\,keV) $^7$Be neutrinos that
are believed to make up roughly 10\% of the total solar neutrino
flux~\cite{std-solar-model}.  However,
the possibility of observing the much less common $pep$ neutrinos
($E = 1.44$\,MeV) is also foreseen~\cite{c11paper}.

Borexino will directly observe only the electrons scattered by these neutrinos,
not the total neutrino energy.  Hence the observed energy spectrum for each
type of neutrino will consist of a nearly flat signal ending at a shoulder,
at 667\,keV for the $^7$Be neutrinos and 1.22\,MeV for the $pep$ neutrinos.
For this reason, the two important
energy regions for the detector are the main neutrino window (NW) of
250--800\,keV, and the $pep$ window from 800--1300\,keV.  The main window
has a lower limit of 250\,keV due to the unavoidable presence of $^{14}$C, a
$Q = 156$\,keV $\beta$-emitter, in organic material.  The chosen windows
extend beyond the nominal spectra endpoints due to the finite energy resolution
of the detector.  The Borexino experiment is made challenging by the fact
that many naturally-occurring radioactive isotopes can decay to produce
signals in these energy windows.

\begin{figure}[h!]
\begin{center}
\epsfig{file=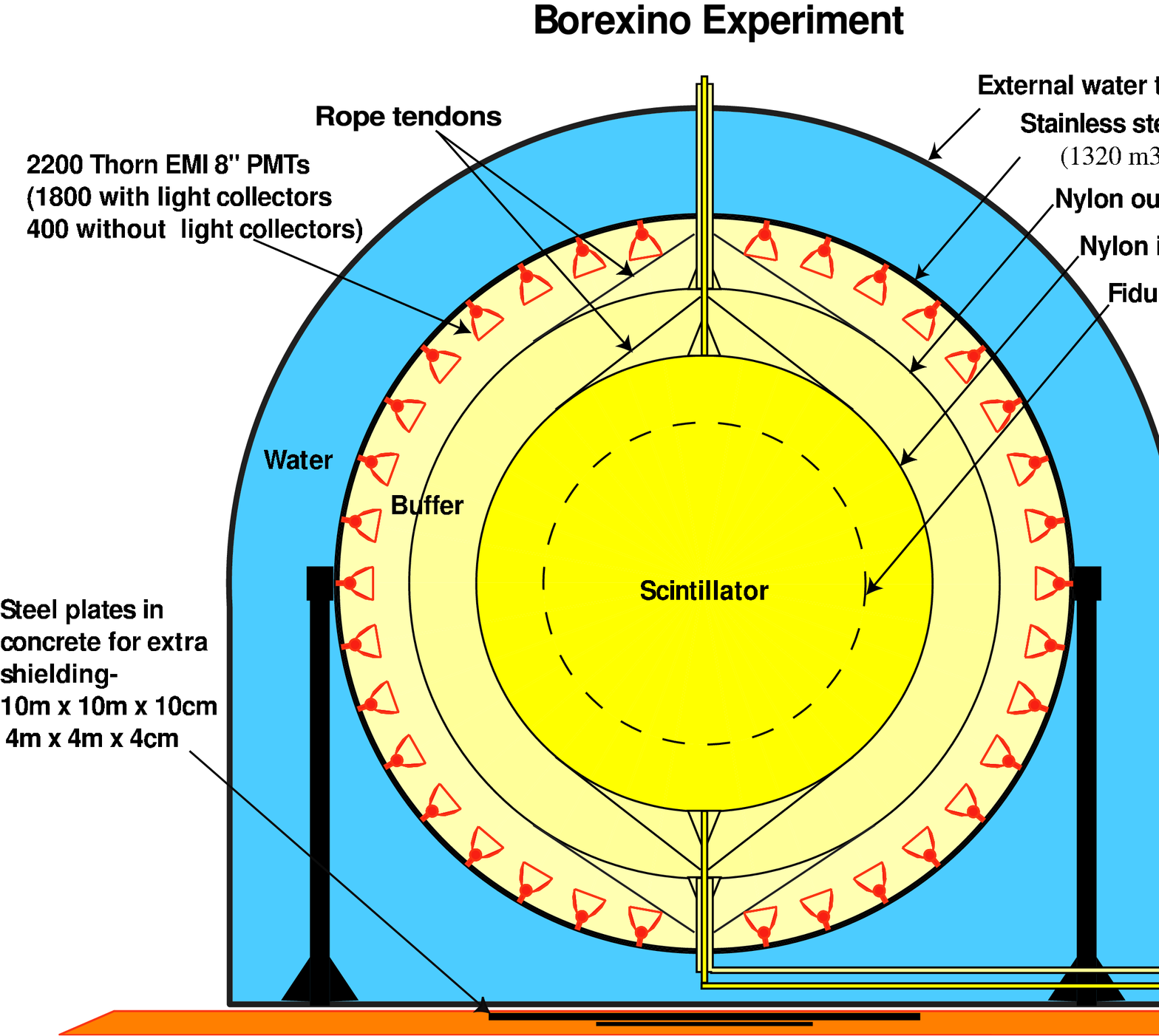, height=4in}
\caption{Schematic diagram of the Borexino detector in cross-section.}
\label{f:borexino}
\end{center}
\end{figure}

The active component of the Borexino detector (Figure~\ref{f:borexino}) is a
spherical shielded mass of slightly under 300~tons of liquid scintillator.
When a neutrino scatters from an electron in the scintillator, the recoil of
the electron induces light emission in the material.  The main component of the
scintillator is pseudocumene, an aromatic solvent (1,2,4 trimethyl benzene)
that emits light in the ultraviolet at $\sim 280\,$nm.  The fluorescent dye PPO
(2,5-diphenyl oxazole) was selected~\cite{pc+ppo} as a pseudocumene additive at
1.5\,g/liter.  Energy is transferred in a fast, non-photonic manner from
pseudocumene to PPO molecules.  PPO acts as a wavelength shifter: because it
has a large Stokes shift between its absorption (290\,nm) and emission
(380\,nm) wavelengths, it greatly reduces re-absorption of the emitted light,
lengthening the optical attenuation length to $\sim$7\,m.  Scintillation light
is detected by an array of 2240 eight-inch, model ETL 9351 photomultiplier
detectors~\cite{etl9351}, most equipped with light
collectors~\cite{prochaska,light-cones} to enhance their optical coverage.  The
phototubes are fixed to the inner surface of a 13.7-m diameter stainless steel
sphere centered on the active portion of the detector.

The design of the Borexino detector is based on the principle of graded
shielding: traveling inward to the center, each component is protected from
external radiation by the preceding one.  To reduce external radioactive
background (mainly inward-bound $\gamma$ rays from various outer portions of
the detector) contaminating the data sample of neutrino-induced scintillation
events, data will be analyzed primarily from the central scintillator volume of
diameter 6\,m (having a mass of 100~tons), considered to be the fiducial volume
of the detector.\footnote{A scintillation event may be determined to have
occurred within or outside the fiducial volume via methods of position
reconstruction that use timing information from the
phototubes~\cite{posn-recon1,posn-recon2}.}  The expected rate of $^7$Be
neutrino events in the main neutrino window within this fiducial volume is
$\sim30$ per day.  (A few additional events per day from $pep$ and CNO
neutrinos will also be seen in this energy window.)  The size of the fiducial
volume may be changed as necessary by trivial adjustments to data analysis
software.  All 300~tons of scintillator are contained by a transparent
spherical nylon vessel, the so-called inner vessel, with an 8.5\,m diameter.  

A passive buffer region outside the inner vessel shields the scintillator from
the radioactivity of the photomultiplier tubes and the stainless steel sphere.
If the scintillator were in contact with these components, which are relatively
high in radioactivity, the rate of events would overwhelm the data acquisition
system.  The buffer fluid consists of pseudocumene with an added component
(dimethyl phthalate or DMP~\cite{bx-quenching} at 5~g/liter) that quenches
scintillation: the optical attenuation length for scintillation events in the
buffer is only about 20\,cm.  Events in the buffer are therefore detected by no
more than a few photomultiplier tubes, and are easily discriminated from
neutrino events that occur in the scintillator fluid.  We note that the pseudocumene and DMP
mixture has nearly the same density as the scintillator~\cite{pocar}.  Thus
the choice of scintillator and buffer materials results in a near
\emph{buoyancy-free} environment for the scintillator, permitting the use of
a thin membrane for the nylon vessel.

A second nylon vessel of diameter 11\,m, the outer vessel, divides the
buffer fluid into inner and outer regions in order to prevent radioactive
impurities (radon, dust) from approaching the inner vessel.  The volume of
buffer fluid outside the outer vessel is contained by the stainless steel
sphere.  Beyond the stainless steel sphere is an outer steel tank filled with
an ultra-pure water buffer; this is used as an active muon veto system and as a
passive shield against neutrons from the rock walls of the laboratory.

The Borexino detector is extremely sensitive to backgrounds from trace
quantities of radioactive impurities that occur in dust and within detector
materials ($^{238}$U and $^{232}$Th decay chains and $^{40}$K) and in air
($^{39}$Ar, $^{85}$Kr, and $^{222}$Rn and its daughters).  In addition to
providing methods for removing such impurities from the
scintillator, great care must be exercised to avoid
contamination during fabrication and handling of critical components of the
detector.  In particular, because of their close proximity to the sensitive
part of the detector, the nylon vessels require careful selection of materials
and clean procedures for fabrication and handling.  The vessels are also rather
delicate and must be maintained within particular
ranges of humidity, temperature, and differential pressure in which the
membrane stress levels are acceptable.

\section{The Design Requirements for the Vessels}
\label{s:requirements}
We first must point out that the choice to use two flexible, thin nylon vessels
in the Borexino experiment is not obviously the only feasible option.
Previous neutrino detectors, both \v{C}erenkov ({\it e.~g.}, SNO~\cite{SNO}) and
scintillator-based ({\it e.~g.}, CHOOZ~\cite{CHOOZ} and Palo
Verde~\cite{PaloVerde}), have used a single rigid, transparent,
thick-walled acrylic vessel for liquid containment.  A rigid vessel has several
advantages: it can be self-supporting (no complex support structure is
required); it has a fixed, well-known volume, minimizing systematic
uncertainties in the target mass; and because of its rigidity, precise
control over the internal and external liquid levels is not necessary
during initial filling of the detector.

Despite these advantages, other considerations made it difficult to choose a
rigid vessel design.  Common rigid plastics such as acrylic and polycarbonate
are not chemically compatible with the pseudocumene scintillator.
Even more importantly in our case, the low rate of radioactive background
tolerable in Borexino causes the thick walls of a rigid vessel to become a
liability.  Since solid materials cannot be purified to the same standards as
the liquid scintillator or the passive buffer, the scintillator containment
vessel, if it is too massive, can contribute disproportionately to the number
of radioactive background events (particularly $\gamma$ rays) seen in the
fiducial volume of the detector.  This led us to opt for a thin-membrane
design, which has the added benefit of greater optical clarity.
Thin-membrane vessels also have the
advantage of being possible to construct in a controlled (clean room)
environment, unlike a rigid acrylic vessel, which would need to be built
on-site.

Once thin membranes are selected, a number of other design constraints appear.
Not only does an inner vessel (IV) contain the scintillator, but in addition a
second outer vessel (OV) surrounds it.  The OV divides the buffer fluid into
two concentric volumes, in order to act as a barrier toward radon gas and to
keep any particulate remaining within the Stainless Steel Sphere well away from
the scintillator fluid.  However, this nesting of two vessels complicates the
design further.  Systems are needed to keep the vessels stationary under the
influence of buoyant or gravitational forces, due to potentially different
densities (resulting from composition or temperature differences) in the three
separate fluid volumes.  These are provided in Borexino by sets of ropes
enveloping the vessels like hot-air balloons.  Rigid nylon rings, or ``end
caps,'' at the polar regions (top and bottom of the detector), two for each
vessel, serve as
fixed attachment points for the vessel membranes.  Tubes passing through the
end caps allow insertion and removal of fluids.  The end caps and tube
assemblies also act as sites for attachment of monitoring instruments and the
ends of the rope systems.

As opposed to a rigid vessel design, the vessels themselves cannot support
significant weight or buoyant forces.  All such forces are transferred, through
the system of ropes and end caps, into the external stainless steel sphere
(SSS); the vessels and sphere work as an integrated design. The region outside
the SSS is a buffer of ultra-pure water, so it is necessary to ensure that the
lower-density volume contained by the SSS does not float upward.  Engineering
of the sphere to handle these loads is straightforward: the SSS is affixed by
several legs and pads to the bottom of the outermost steel tank.  The sphere
doubles as a support mechanism for all the internal phototubes.

\begin{figure}[h!]
\begin{center}
\epsfig{file=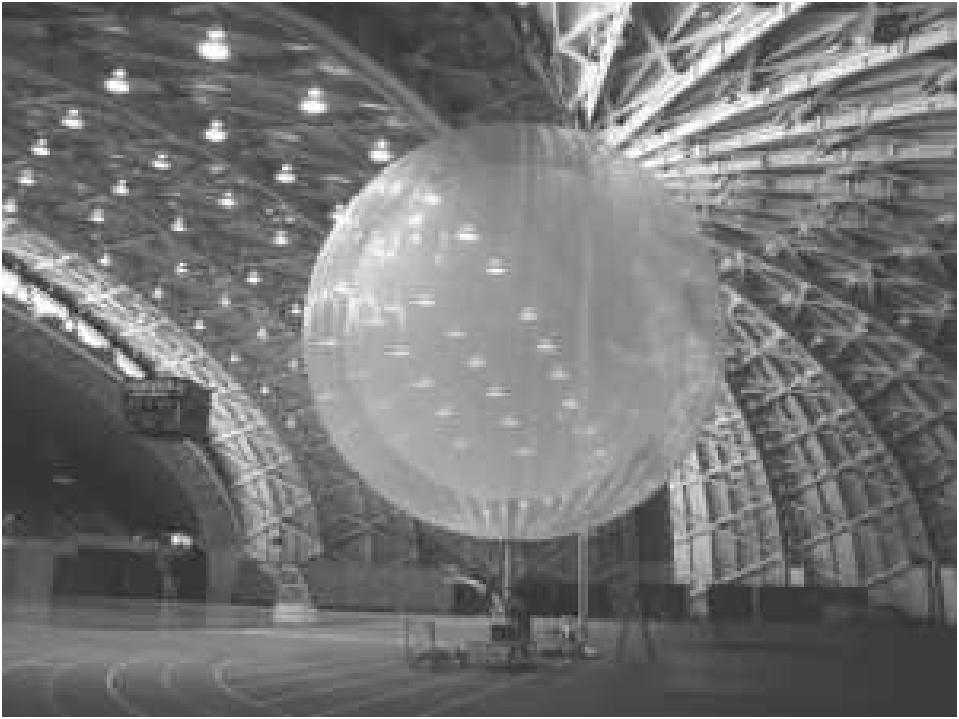, width=5in}
\caption{Prototypes of the nested nylon vessels.  They were built in a
fashion similar to that of the real vessels, although not under clean
conditions.  The test inflation shown here was done in the Jadwin
Gymnasium at Princeton University.}
\label{f:proto-vessels}
\end{center}
\end{figure}

Although the thin-membrane vessels of Borexino were not installed into the
detector and inflated to their final spherical shapes until spring 2003, they
have been an integral part of the Borexino experimental design since the
initial proposal was submitted in 1991~\cite{bx-proposal}.  A 4-ton
prototype of the thin-membrane design, the Counting Test Facility, has been
operated successfully since 1995~\cite{ctf}.  A
full-scale mock-up of the nested Borexino vessels has also been constructed
and inflated at Princeton University~\cite{pocar,cadonati,pppl-news};
see Figure~\ref{f:proto-vessels}.  Since the
initial Borexino proposal, the concept
has been incorporated in other full-scale experiments as well.  For instance,
the reactor neutrino detector KamLAND uses a thin-membrane scintillator
containment vessel~\cite{kamland-design}.  Although it has no second vessel
analogous to the Borexino OV for restricting the inward flow of
radon, KamLAND does have thin membranes in front of its photomultiplier tubes
to act as a radon barrier.

The fundamental requirements of the Borexino experiment for the IV and OV
are summarized below.  The vessels must be able to survive their environments:

\begin{itemize}

\item[a)] {\bf Chemical resistance.}  
The vessels must be chemically compatible with the scintillator (pseudocumene
and PPO) and buffer (pseudocumene and DMP); with pure water; and of course
with normal air, in which they were constructed and with which they were
initially inflated. 
For Borexino, the selected filling strategy involved filling the inner detector 
({\itshape i.~e.}, everything inside the SSS) first with water, then later with
the scintillator and buffer fluids. 

\item[b)] {\bf Mechanical strength.} The vessels must withstand the expected 
membrane stresses due to buoyant forces that occur during filling and 
steady-state operations.  By design they should also be able to withstand, at
least for long enough to correct the situation, stresses
up to 20\,MPa that could occur as a result of 5$^\circ$C temperature
differences (which would cause 0.4\% density differences) between any
of the three fluid volumes.  The expected stresses
are calculated by a finite element analysis, discussed below in
section~\ref{ss:stress}.  Finally,
they must survive repeated handling during installation and inflation
operations that could lead to crack formation (brittle failure mode).

\end{itemize}

The vessels must in addition not hinder the operation of the experiment:

\begin{itemize}

\item[c)] {\bf Optical transparency.} 
The vessels must be transparent and free of haze for the blue and near-UV
light emitted by the scintillator (350--450\,nm).  They must also have an
index of refraction similar to that of pseudocumene.  This minimizes refraction
of scintillation light at the vessel and reflection of light from the
nylon-scintillator interface, both of which could interfere with an accurate
position reconstruction of scintillation events.
The extents to which nylon films meet this requirement and the two previous
ones are described in section~\ref{ss:film-selection} below.

\item[d)] {\bf Low intrinsic radioactivity.} 
The levels of U, Th, and K in the scintillator vessel must be low to minimize
background due to gamma rays that originate in the vessel.  (Potassium is an
issue due to the long-lived, naturally occurring radioactive isotope $^{40}$K,
which occurs in natural potassium with an isotopic abundance of 120\,ppm.) In
addition, emanation of $^{222}$Rn gas due to intrinsic $^{226}$Ra in the nylon
must be low.  The desired requirement is that the vessels contribute fewer than
one radioactive event per day in the $^7$Be neutrino energy window within the
fiducial volume.  The intrinsic background activities in the nylon film are
described towards the end of section~\ref{ss:film-selection}, while the
repercussions for Borexino are discussed further in
section~\ref{sss:rn-emanation}.

\item[e)] {\bf Clean fabrication.} Fabrication of the vessels (cutting of panel
sections, bonding of joints, packaging, etc.) must be done in clean conditions,
as we describe in section~\ref{s:design-fabrication},
to minimize backgrounds due to dust: the tolerable amount of dust within
the IV is no more than 3\,mg.  Also, deposition of radon daughters on
the vessel surfaces must be kept to an absolute minimum.
It is worth noting that even before the detector was operational, for instance
during the transportation and installation of the
Borexino vessels at LNGS, the OV was helping to protect the IV from dust and
radon exposure.

\end{itemize}

Finally, the vessels must be effective at fulfilling the goals for which they were designed:

\begin{itemize}

\item[f)] {\bf Low permeability.}
To minimize backgrounds due to diffusion of $^{222}$Rn from the buffer through
the vessel into the scintillator, the membrane must have a low permeability to
radon.  In brief, the film must be impermeable enough (and thick enough) that
the average time required for a radon atom to diffuse through the entire film
is comparable to or greater than the mean life of radon, 5.516~days.  The
permeability of nylon to radon atoms is described in
section~\ref{sss:nylon-permeability}.

\item[g)] {\bf Leak tightness.} The vessels, the IV in particular, must be leak
tight enough to prevent any significant mixing between the scintillator and the
buffer fluids during the lifetime of the experiment.  We state the leak
tightness requirements in section~\ref{ss:leak-tightness}, and report
measured values there and in section~\ref{ss:leak-rate-tests}.

\item[h)] {\bf Monitoring.} The vessels must be outfitted with instrumentation
that permits monitoring the temperatures and pressures of the scintillator and
buffer fluids, as well as the current shapes and positions of the vessels
themselves.  The monitoring instrumentation is described in
section~\ref{ss:instrumentation}.

\end{itemize}

Note that requirements a), b), c), d) and f) above are mainly or entirely a
function of the material chosen for the vessels, rather than the method of
fabrication or the supporting infrastructure.

\section{Production and Selection of Nylon Film}
\label{s:nylon-film}

Thin nylon film can meet all requirements for the vessels and was thus the chosen
material.  However, standard commercially available films are not produced under
sufficiently clean conditions.  To achieve the desired properties and level of
cleanliness, the film was therefore extruded under appropriate conditions and
from raw materials that were carefully selected such that the material meets
the size, purity, optical clarity, and cleanliness required for Borexino.  Some
details of the properties of the materials and of the vessel fabrication
sequence are summarized below and in the following section.

\subsection{Production of the nylon film}
\label{ss:production}

The term nylon, as described in great detail in~\cite{nylon} (for instance),
refers to a family of polymers built from carboxylic acid, amine, and/or amino
acid monomers.  Nylons may be characterized in numerous ways, the most obvious
being by the chemical formula of the monomers (the candidate materials for
Borexino were all based on nylon-6, H-[HN(CH$_2$)$_5$CO]$_n$-OH).  To produce a
thin nylon film, polymer pellets are heated above their melting point (to
around 250$^\circ$C), yielding a melt that is extruded at high pressure.  Rapid
cooling afterwards ensures that the polymer chains remain in a transparent
amorphous state rather than developing a hazy crystalline structure.

Several types of pellets were initially under consideration for use in the
Borexino vessel material.  The two final candidates, selected mainly due to
radiopurity considerations, included Capron B73ZP pellets made by
AlliedSignal/Honeywell~\cite{honeywell}, and Sniamid ADS40T pellets
manufactured by Nyltech.  (Since then, the Capron product line has been
acquired by BASF~\cite{basf}, and the Sniamid line by
Rhodia Engineering Plastics~\cite{rhodia}.)  Capron B73ZP
consists of pure nylon-6 polymer chains, while
Sniamid ADS40T is a nylon-6 based co-polymer ({\it i.~e.}, its polymer chains
contain more than one type of monomer) with a proprietary formula.

Nylon film produced from Sniamid pellets proved to be slightly brittle,
however; an additive was required for pliability.  The selected additive was
Ultramid~B4, another pure nylon-6 polymer manufactured by BASF~\cite{basf}.
Sniamid and Ultramid pellets were mixed in a 5:1 ratio, upon extrusion
yielding a film that will also be referred to as Sniamid.

\subsubsection{Radiopurity levels of the nylon pellets}

We set a radiopurity target of 1 part per trillion (ppt) by mass of uranium in
the pellets and the film for the inner vessel; anything less than or equal to
5\,ppt was also considered acceptable, since it would not affect Borexino's
sensitivity to $^7$Be solar neutrinos.  Efforts were made to measure the
$^{238}$U in the pellets which are extruded to make the film. These
measurements were carried out by Tama Chemicals~\cite{tama} with inductively
coupled mass spectroscopy on samples of nylon digested in ultrapure acids.

The Capron B73ZP was measured in this way to have contamination levels of
0.46\,ppt U by mass and 1.1\,ppt~Th~\cite{lowrad}.
Tama Chemicals found the
Sniamid pellet levels to be 1.1\,ppt~U and 1.6\,ppt~Th.  (Errors were all $<
10\%$.)  At this point, therefore, the Capron B73ZP pellets appeared to have a
slight advantage.  Measurements of the pellet potassium contamination (all
isotopes) using neutron activation analysis and graphite furnace atomic
absorption spectroscopy
at various facilities gave inconsistent results, in the range 13--25\,ppb for
Capron pellets and 1.6--25\,ppb for Sniamid pellets~\cite{lowrad}.
All of these measurements were difficult and subject to contamination.
Fortunately, the most important values, the radon emanation rates of the final
extruded films, could be measured directly, and were found to be
satisfactory (Section~\ref{sss:film-radioactivity}).

The Ultramid pellet
contaminant levels are not directly known, since the need to include this
additional component in the Sniamid-based film was not realized until after the
pellet radiopurity measurement campaign.  In any case, the Sniamid film was
thought of as a backup material at this point.

\subsubsection{The nylon film extrusion process}
\label{sss:extrusion}

The process of extrusion takes nylon pellets in raw form and converts them into
flat sheets.  The thickness requirements for the Borexino vessels are chosen as
a compromise between increased thickness, for mechanical strength and reduction
of radon diffusion; and reduced thickness, for flexibility (which minimizes the
likelihood of cracking) and reduction of the total radioactive decay rate
by minimization of the vessel masses.  In addition, we needed sheets that were
as wide as possible, in order to minimize the number of panels from which
each vessel would have to be constructed.  Sheets with
a thickness of 125\,$\mu$m and width of about 122\,cm (4\,ft.) were selected.
The thickness was measured during extrusion and has a tolerance of
$\sim5\,\mu$m. It should be noted that the thickness is greater than that of
typical commercially produced, food-grade nylon film.  The width of 4\,ft.\ is
more or less the maximum available on the market.

The candidate nylon films (``Capron,'' generated from pure Capron B73ZP
pellets, and ``Sniamid,'' produced from the Sniamid ADS40T pellets and Ultramid
B4 pellets in a 5:1 ratio) were extruded at two separate facilities.  Capron
film was produced at an AlliedSignal/Honeywell plant in Pottsville, Pennsylvania, USA,
in 2001.  Sniamid film was extruded at the mf-folien plant~\cite{mf-folien} in
Kempten, Germany, later that year.  The two plants are among the few possessing an extrusion apparatus
suited for producing wide panels of such thickness under relatively clean
conditions.  The temperature and rate of extrusion at both facilities were
closely monitored and optimized to ensure the crucial property of low haze
levels ($< 1.5\%$ scattering of incident light) in the newly extruded nylon
film.

\subsubsection{Pre-cleaning of the nylon film}
\label{sss:precleaning}
 
The inner surface of the inner vessel must satisfy very strict particulate
contamination standards due to the relatively high levels of U, Th, and K in
dust. Particles deposited on the nylon film could end up in the scintillator
fluid, contributing high levels of background in the detector. It was estimated
that, assuming 1\,ppm U and Th concentrations in generic particulate, a surface
cleanliness level of 50 as defined by US military standard 1246-C
(at most one particle with a 50\,$\mu$m diameter per
square foot \cite{milstd1246c}) is required. In other words, under these
assumptions, only 3\,mg of dust are tolerable on the inside surface of the
entire inner vessel.

All nylon film used for both inner and outer vessels was precision cleaned
and certified at level 25 by CleanFilm Inc., New York, USA.  Cleaning used a
non-contact, ultrasonic disruption technique to loosen particles that were then
removed from the surface by negative pressure.  The cleaned nylon rolls were
double-bagged with thin, commercially available class-25 nylon film, and
covered with an aluminized layer to minimize inward radon diffusion during
shipping and storage.  The first few external layers of each roll were
discarded during fabrication.  All film used to cover the vessel panels during
vessel assembly and shipping was also certified to level 25 (refer to
section~\ref{ss:fabrication}).

\subsection{Selection of the nylon film}
\label{ss:film-selection}

Several factors were considered in the selection of the particular nylon
material for the vessel, many of which, such as low radon emanation,  are not
standard tabulated data.  A program of specific measurements was therefore
carried out to characterize and select the final envelope material.

\subsubsection{Chemical compatibility}

The mechanical properties of dry nylon film are thought to be essentially
unchanged after immersion in pure water-free pseudocumene, scintillator fluid
(pseudocumene with added PPO), and buffer fluid (pseudocumene with added DMP).
Tests of the material properties (tensile strength $\sigma_t$, and Young's
modulus $E$) of Capron film immersed in pure pseudocumene, in equilibrium with
air at relative humidities varying from 0--60\%, show essentially no difference
from Capron film in air at the same relative humidities~\cite{mccarty}.
Although an earlier program of tests showed some variation in the material
properties of nylon immersed in pure pseudocumene, scintillator, or buffer
fluid compared
with nylon in a dry N$_2$ atmosphere~\cite{cadonati}, it is believed that this
variation resulted from traces of water present in the respective liquids.
Samples of various nylon films immersed in any of these liquids that were kept
in a tightly sealed jar for three months, or in a liquid to which silica gel
was added (to absorb water), showed no significant difference in their tensile
strength or Young's modulus from samples in a dry nitrogen
atmosphere~\cite{cadonati}.

An important concern is the effect of water on the nylon film over periods of
several months, as in the water-filling stage of Borexino.  The Young's modulus
of nylon in contact with water (even when immersed in another fluid that is
saturated with water) is reduced due to plasticizing---hydrogen bonds form
between water molecules and polymer chains in place of hydrogen bond crosslinks
between chains.  However, the effect is found to be reversible, even after the
film is soaked in room-temperature water for one month~\cite{cadonati}.  Tests
have shown that nylon films come to equilibrium with the relative humidity of
the surrounding environment within a few days.

Nylon film that has been soaked in water and then allowed to dry again exhibits
a milky white haze on its surface.  This is thought to be a layer of nylon
monomers leached from the material by the water and then deposited on the
surface.  The monomer is initially present throughout the nylon; it
diffuses over time, especially when the nylon is humidified.  This layer
of monomer can be washed away by another immersion in water or even in
pseudocumene.

Water can also cause hydrolysis in nylon.  This is the reverse of the
polymerization reaction, compromising the nylon molecular structure, but it
only happens on time scales that are long compared to the expected period of
several months for the Borexino water filling.  In particular, the Young's
modulus of a nylon-66 sample, chemically similar to the nylon-6 based materials
used in Borexino, is drastically reduced when it is maintained in a 100\%
relative-humidity environment for 2~months at
66$^\circ$C~\cite{nylon-hydrolysis}.  An extrapolation from data at 66$^\circ$,
82$^\circ$ and 93$^\circ$~\cite{nylon-hydrolysis} to the much lower operational
temperature in Borexino ($\sim15^\circ$C) gives an estimated time scale of
10~years for significant degradation to occur~\cite{cadonati}.
Thermodynamic arguments yield similar time frames~\cite{nylon-properties}.

\subsubsection{Mechanical measurements}

Polymer materials in general have a glass transition temperature $T_g$, at
which a second-order phase transition occurs, that depends upon the exact
composition of the material~\cite{rodriguez}.  Above $T_g$, the material is in
a ``plastic state'' which is pliable and, if stretched, may be permanently
deformed.  At lower temperatures, a ``glassy state'' has a greater Young's
modulus, but is more brittle and breaks at lower strain.  Notably, the presence
of water (higher relative humidity) decreases the value of $T_g$ from that for
a dry polymer.  Hence at constant temperature between the minimum and maximum
possible values for $T_g$, nylon film in a humid environment is likely in the
plastic state, while dry nylon is in the glassy state.  During construction and
installation of the Borexino vessels, a high humidity was maintained to prevent
brittle failure.  Once the vessels are filled with scintillator and buffer
fluids and are being maintained in a static condition in their preferred
spherical shape, the humidity of their environment will be lowered in order to
take advantage of the higher tensile strength and Young's modulus.

\begin{figure}[h!]
\begin{center}
\epsfig{file=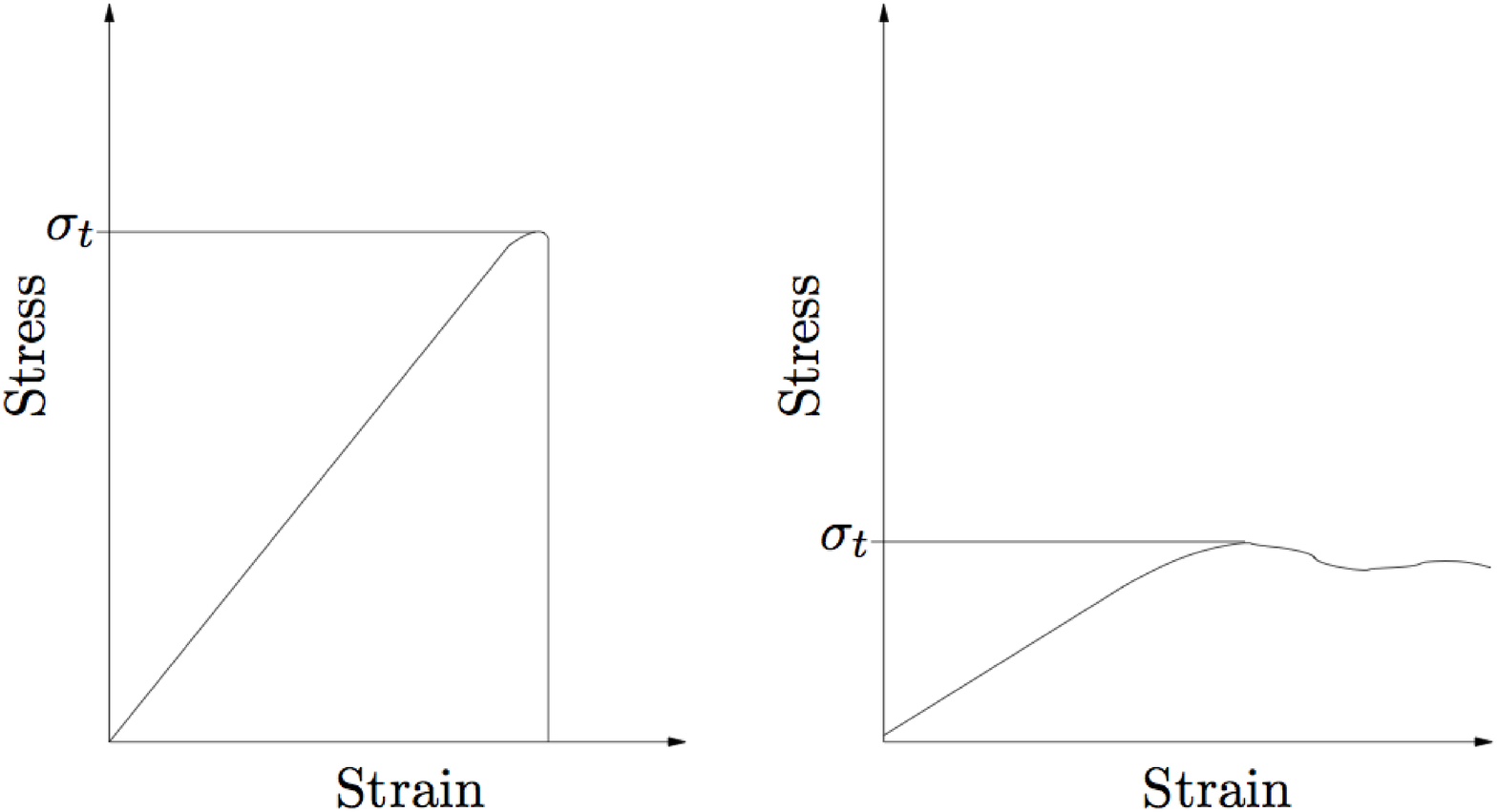, width=4in}
\caption{Schematic diagrams (not to scale) of stress-strain relationships for
nylon films below the glass transition temperature (left) and above it (right).
Typical values for the tensile strength $\sigma_t$ are 75\,MPa for film in
the glassy state, and 20\,MPa for film in the plastic state.}
\label{f:stress-vs-strain}
\end{center}
\end{figure}

A series of mechanical tests was performed on Capron and Sniamid film, in order
to measure their material
properties as functions of relative humidity, holding the temperature constant
at 22$^\circ$C~\cite{mccarty}.  For each individual sample, a Tinius-Olsen
machine was used to graph the stress $\sigma$ on the sample as a function of
its fractional elongation (strain) $\epsilon$.
Typical stress-strain relationships for the film are shown in
Figure~\ref{f:stress-vs-strain}.  The stress is approximately linear with
strain, $\sigma = E \epsilon$, and the film behaves elastically according to
Hooke's Law nearly up to the yield point.  At stresses beyond the yield point,
the material is irreversibly damaged.  Film in the plastic state, above
$T_g$, undergoes viscous deformation, elongating to several times its original
length.  Film in the glassy state, below $T_g$, elongates much less and then
breaks.

The Young's modulus values at room temperature are 1.7\,GPa when the
film is dry, and 0.4\,GPa for wet film at 100\% relative
humidity~\cite{mccarty}.  With 125\,$\mu$m-thick film, typical values at room
temperature for the membrane stress at the yield point are 70--80\,MPa and
$\sim 20$\,MPa for dry and wet conditions, respectively.  These values are
similar for Capron and Sniamid films, although Sniamid is a little
stronger at all humidities and appears to transition from the glassy to the
plastic state at a slightly higher relative humidity.

Another factor to consider is creep, an irreversible stretching of the film
when a constant stress is continually applied over a long period of time.  At
room temperature, creep at a stress level of 5\,MPa for typical nylon films is
0.5\%, 3\%, and 7\% under dry, 20\% relative humidity, and wet conditions,
respectively~\cite{cadonati}.  The
Borexino design will maintain stress levels of $<$ 10\% of the yield stress.
As small creep-induced elongations occur at localized stress hot spots,
operational stress levels will naturally be minimized.

The other possible failure mode of nylon film is brittle failure, in which
nylon that is folded back upon itself, forming a crease or point, develops a
crack or pinhole.  Only the glassy state of the film is susceptible to
cracking, the plastic state being much more pliable.  This was nevertheless a
concern because large cracks developed in the nylon film of the second Borexino
prototype, the Counting Test Facility (CTF~2), while it was exposed to a dry
nitrogen atmosphere.  Because of this event, further tests were performed on
the nylon films to simulate the rough treatment and possible brittle failures
that could happen during shipment and inflation of the Borexino vessels.
Inflatable nylon packets made of two 28\,cm circles of film sealed at the edges
were produced.  Packets at different humidities were repeatedly inflated and
deflated.  The deflation step created severe wrinkles in the film and joint
that reappeared at every cycle in the same spots.  The packets were tested for
leaks after each cycle.  Dry packets typically failed after 20 cycles; the time
to failure more than doubled at 40\% relative humidity.
Capron film packets tended to survive
longer than those constructed from Sniamid~\cite{mccarty}.  In any event, the
Borexino film, being only 25\% the thickness of that used in the Counting Test
Facility, should be much less susceptible to cracking.

\subsubsection{Optical measurements}

Among other reasons, one good argument in favor of the use of nylon film for
the scintillator containment vessels is that the index of refraction of nylon-6
at near-UV wavelengths is 1.53.
Since this is so near the value of $n = 1.50$ for pseudocumene,
reflection and refraction of scintillation light from the vessels will be a
negligible problem.

Nylon film can be produced with excellent light transmittance and low haze
properties.  Such characteristics minimize scattering and loss of photons as
they travel to the photomultiplier tubes.
These conditions are essential for accurate energy
measurement, good energy resolution, and precise position reconstruction of the
events.

Capron and other non-amorphous nylon polymers tend to be
slightly hazy at the 0.1\,mm thickness required for the vessel.  This issue
was resolved by rapid quenching of the Capron film during its extrusion,
preventing the development of crystallization that tends to cause this haziness.
The amorphous co-polymer nylons such as Sniamid are by nature very clear
optically.  The nylon films selected for Borexino both show transmittance
$>90\%$ in pseudocumene above 300\,nm
and less than 1\% scattering at angles greater than
70$^\circ$ with 366\,nm light~\cite{cadonati}.  The Capron film was measured
during extrusion to have an average haze level of 0.5\%, much lower than
the required 1.5\%.

\subsubsection{Radioactivity requirements and measurements}
\label{sss:film-radioactivity}

Part of the duty of the Borexino vessels is to act as barriers to the inward
travel of radon gas.  Diffusion of radon through nylon film should be
low---both to reduce emanation of radon from intrinsic $^{226}$Ra in the
material, and to slow the transport of external contaminants.  As we discuss in
more detail in section~\ref{sss:nylon-permeability}, the permeability of nylon to radon
atoms increases rapidly with humidity; this fact must be balanced against the
risk of brittleness that develops when nylon is dry.  It should be mentioned
that the mobilities of noble gases such as radon in nylon are much higher than
those of the heavy metals such as U and Th.

Emanation is a second concern.  This is the process by which radioisotopes
(mainly radon) that are initially embedded in the nylon film may migrate out,
eventually contaminating the scintillator.  The intrinsic bulk contamination of
the nylon will produce events for the lifetime of the experiment, and so must
be as low as possible.  

\begin{table}
\caption{\label{t:RaInNylon}Measured $^{226}\mbox{Ra}$ in the two nylon film
candidates~\cite{rn-sniamid}.  Although the Capron sample extruded for this
measurement was only 100\,$\mu$m thick, both Borexino vessels are made from
125\,$\mu$m film. }
\begin{center}
\small 
\begin{tabular}{lccc}
\hline\hline
Sample & Thickness & $^{226}\mbox{Ra}$ activity  & $^{238}\mbox{U}$ equivalent  \\
       & [$\mu$m] & [mBq/kg] & [ppt] \\
\hline
Capron  & 100 & $0.21\pm0.03$ & $17 \pm 2 $ \\
Sniamid & 125 & $< 0.021$     & $< 1.7$ \\
\hline
\end{tabular}
\end{center}
\end{table}

To determine the intrinsic contamination of $^{226}$Ra (the radon progenitor)
in the nylon films, our Heidelberg collaborators used a targeted measurement
technique~\cite{rn-sniamid} in which the radon emanation is measured directly
from the film in presence of water (which enhances the emanation).  Ten-kg
film samples were extruded and carefully cleaned, in the same conditions as the
film used for the vessel construction, then loosely rolled and placed into an
emanation chamber.  The radon emanated from the film was collected and counted
by low background detectors.  The results~\cite{rn-sniamid} are listed in
Table~\ref{t:RaInNylon}: the Capron film was measured at
$210 \pm 30$\,$\mu$Bq/kg of $^{226}$Ra activity, while only an upper limit was
observed for the Sniamid, at 21\,$\mu$Bq/kg.

Since the contamination level in Sniamid film was better than that in Capron by
a factor of ten, Sniamid was decisively chosen to be the inner vessel material.
The resulting rate of events in the fiducial volume of Borexino would range, in
the worst-case assumption of rapid and complete scintillator mixing, between 1
and 7 events per day per 100 tons of scintillator~\cite{rn-humidity}, depending
upon the water content of the scintillator in contact with the inner vessel.
(The issue of radon emanation from nylon is discussed further in
section~\ref{sss:rn-emanation}.)
If the intrinsic event rates of $^{226}$Ra decays are in secular equilibrium
with the isotope $^{238}$U at the top of the decay chain (not necessarily a
safe assumption, as equilibrium may be broken by chemical processes), the
inferred bulk contamination levels of uranium in the nylon are 17\,ppt by mass
in the Capron and $<$ 1.7\,ppt in the Sniamid.  Comparison of these numbers
with those reported for the raw nylon pellets (section~\ref{ss:production})
suggests that the Capron film was slightly contaminated during the extrusion
process, while the Sniamid was not; or, alternatively, that the equilibrium
between $^{238}$U and $^{226}$Ra was indeed broken.

The figures above exclude a surface component of radon emanation, likely due to
dust adhering to the film.  This component made up roughly 40\% of the
intrinsic bulk contamination in the case of one sample not subjected to
pre-cleaning.  On samples that had been pre-cleaned, the surface component was
negligible, justifying the pre-cleaning procedure described earlier.

\subsubsection{Film selection summary}

Capron and Sniamid pellets displayed the lowest U and Th radioactivity levels
among the set of candidate materials.  Once extruded into films, both had
excellent mechanical and optical properties.  Capron had a slight overall edge
over Sniamid mechanically; although it exhibits a slightly lower yield point
and Young's modulus, it performed better in the packet tests. It was thus
chosen as the prime candidate (the two films performed similarly above 30\%
relative humidity).
Only after the outer vessel was already assembled did radon emanation data
from both films became available. Sniamid was more than an order of magnitude
better than Capron~\cite{rn-sniamid}, and was therefore used
to build the inner vessel.

\section{Fabrication of the Nylon Vessels}
\label{s:design-fabrication}

The main challenge in the design of the Borexino vessels is to achieve low
contamination, both in the nylon film and in the bulk materials of the polar
regions.  An important factor to consider is the exposure of the nylon film to
ambient radon and dust particles during fabrication of the vessels.
 
The vessels were fabricated in a Class 100 clean room using a method in which
panels are bonded together on a table.  Assembly in a clean room guarantees
very low particulate contamination. In addition, a newly designed radon filter
had been inserted along the makeup air-line that feeds the vessel construction
clean room, allowing us to reduce the radon exposure of the vessel surfaces.

The film, originally extruded from nylon pellets, was first pre-cleaned with
electrostatic techniques, then cut into sheets and bonded together to form two
concentric nested spheres.  The inner vessel (IV) is made of 36 panels while
the outer vessel (OV) is made of 40 panels.  These individual panels were
joined together at seams to form the spherical vessel shapes via a
semi-automated bonding method.

Once the envelopes were fabricated, circular nylon end plates were inserted at
each pole of both vessels. They provided the necessary transition to join the
nylon film and the tubes carrying fluid into the vessels. With the vessel
membranes, they had to satisfy stringent leak tightness requirements.  After
connection to the rest of the tube assemblies, they also serve as structural
pieces that transfer the load from the vessels to the stainless steel sphere
(SSS).  With the polar hardware in place, hold-down ropes, followed by
optical fibers, were installed on each vessel. The IV was then nested inside
the OV, and the final seam of the OV was sealed. The
vessels were finally packaged and shipped to Gran Sasso, where they were
installed within the SSS and inflated with synthetic air
having an ultra-low radon content.

\subsection{Clean room design and control of radon exposure}
\label{ss:cleanroom}

The Princeton clean room for nylon vessel fabrication, manufactured by Control
Solutions Inc., is approximately $22\times6.5\times4.5$ meters in size. The
clean room is certified to class 100, and was measured to be around class 10
when unperturbed.  The class of a clean room is the number of particles
with diameter $\geq 5\,\mu$m per cubic foot of air.  Recirculation time of the air
through HEPA filters is about 30 seconds.  The relative humidity and
temperature were maintained at 50$\%$ and 17 $^{\rm o}$C respectively in order
to keep the nylon film supple and easy to handle.  In order to avoid radon
contamination, the clean room water supply, including that used to maintain
humidity, was aged roughly 100 days.

The clean room is meant to remain filled with low-radon air supplied by a radon
filter.  This requires the clean room system to be leak-tight in order to
prevent back-diffusion of radon.  Being leak-tight also helps to reduce the
amount of radon-free air needed to maintain an overpressure in the room.  The
radon filter relies on vacuum swing adsorption, in which two charcoal columns
are alternately fed and regenerated under vacuum. The columns are switched at
regular intervals ($\sim$ half hour).  The filter supplies the clean room with
85\,m$^3$/hr of low-radon (0.3--0.4\,Bq/m$^3$) air.
For reasons which are not entirely understood, however, the average radon
level within the clean room was rather higher, about 1.5--2\,Bq/m$^3$.
Further information about the vessel assembly clean room and the radon filter,
including some hypotheses to explain this discrepancy,
may be found in \cite{pocar,scrubberpaper}.

Every effort was made to minimize radon exposure of the film.  Radon is
naturally present in air at about 30\,Bq/m$^3$.  Though this figure is
significantly reduced in the clean room, surface exposure of Borexino
materials, especially the nylon vessel, is still a big concern.  Daughters from
the residual radon activity in the clean room can stick to the film, either in
the form of individual ions, or attached to particulates and aerosols in the
air.

Two de-ionizing bars were employed to neutralize static charge buildup that
could attract particulate or charged radon daughters onto the nylon film being
unrolled from the spools.  Nylon panels were kept covered with certified class
25 thin nylon film at all times throughout vessel construction; only strips a
few cm wide at the edges of the panels were uncovered for several hours when
glue joints were made. The cover sheets were removed just before sealing each
vessel with the last joint. Both vessels were kept covered with class 25 film
and a layer of aluminized foil while not being actively worked upon.

In order to estimate the surface density of $^{210}$Pb atoms that may have
plated out onto the inner surface of the IV during its fabrication, it is
necessary to have a model for their deposition.  The simplest possibility is to
assume that every radon atom that decayed in the 3.5-m high air column above
the film, during the time periods in which it was exposed, ended up decaying
and falling onto the film.  Because the nylon vessel film was protected from
exposure to air as much as possible, we estimate that the average exposure time
for any individual surface area on the IV was only about one hour~\cite{pocar}.
This would imply a surface density of $^{210}$Pb of 25\,$\mu$Bq/m$^2$.

Recirculation through HEPA filters, however, has recently been
shown~\cite{cabrera} to be highly effective in removing radon daughters from
air.  Radon gas itself is unaffected by filtration, and some build-up of
$^{210}$Po and $^{210}$Pb occurs because of Rn decays in air after it emerges
from the filter.  The volume concentration of these daughters at the work
surface in the clean room depends on the amount of time for decay, and so is
inversely proportional to the flow velocity of the air.  However, the flux of
these daughters (density times velocity) is independent of the flow.  Thus the
effect of the recirculation flow on plate-out depends on the detailed kinetics
of the plate-out at the material surface, and is difficult to predict.  To
study this, we constructed a small mock clean room (dimensions 2\,m $\times$
2\,m $\times$ 3\,m) and spiked it with a known amount of radon, whose daughters
were allowed to plate out on film samples.  In particular, for a recirculation
time in the mock clean room of $\sim$35 seconds (a single HEPA unit of size
2\,ft $\times$ 4\,ft and linear output velocity 90\,ft/min was used), the radon
daughter plate-out rate at 1.5\,m above the floor was found to be roughly 1\% of
what one would calculate for a 1.5-m high still air column using the naive
model described above.

With the mock clean room and vessel construction clean room having roughly equal
recirculation times, the figure of 25\,$\mu$Bq/m$^2$ above should be multiplied
by the experimentally determined factor of 1\%, yielding an estimated
$^{210}$Pb surface contamination of 0.25\,$\mu$Bq/m$^2$ on the IV inner
surface.  In the worst-case scenario of complete desorption of lead atoms into
scintillator, coupled with thorough convective mixing, this would imply only
1.5 $^{210}$Pb decays per day in the entire 100-ton fiducial volume of
Borexino.  Furthermore, much of this activity will be removed before
scintillator is introduced.  Many $^{210}$Pb atoms (lab-scale tests indicate
80--95\% of them) will have dissolved into the ultra-pure water used in the
first phase of vessel filling, which will later be removed.

The factor of 1\% is an empirically determined value, but a discussion of
various models of radon daughter plate-out that could predict this value would
unfortunately be too far afield.  Results of one common model which uses an
``effective deposition velocity''~\cite{leung-lrt-2004} may be used to estimate
a surface density of plated-out $^{210}$Pb of 0.1\,$\mu$Bq/m$^2$, well within
an order of magnitude of the figure we obtained above, and in fact even
smaller.

Finally, it is worth noting that our estimates of the nylon vessels' exposure
to radon daughters, if accurate, imply that efforts to reduce this exposure
were highly effective.  We may compare the figure of
0.25\,$\mu$Bq/m$^2$ estimated above for Borexino with the activity due to
$^{210}$Pb on the surface of the acrylic vessel of the SNO experiment, for
which no special precautions to protect against radon or Rn daughter exposure
were taken during construction.  For the surface area of the SNO vessel ({\it
i.~e.}, most of it) that was in contact with heavy water for many years, which
presumably rinsed off much of the contamination, the observed $^{210}$Pb event
rate is on the order of $210 \pm 60$\,mBq/m$^2$~\cite{aksel}.  A small portion
of the inner surface of the SNO vessel ``chimney,'' used for filling, was not
constantly in contact with D$_2$O.  Here the measured $^{210}$Pb surface
density was five times higher, $1.07 \pm 0.10$\,Bq/m$^2$~\cite{aksel}.
This is roughly $4 \times 10^6$ times the estimated contamination of the
Borexino IV!

\subsection{Fabrication of the nylon envelopes}
\label{ss:fabrication}

After the initial research and development phase, the nylon vessels were built
in one year with a team of roughly 20 people.  The two vessels were
constructed from wedge-shaped 1\,m wide longitudinal panels.  Adjacent panels
were bonded along their edges and folded into a stack.  Filling lines, support
structures and instrumentation were later attached at the north and south
poles.

Panels were initially allowed to reach equilibrium with the clean room humidity levels.  As nylon absorbs water, it expands by a few percent.  The panels
were then cut to the appropriate shape. The tolerance in the size of
each panel is a few mm, which translates in an uncertainty on the radius of
each vessel of a few cm.  (IV panels are each roughly 13.3\,m
by 0.75\,m; the OV panels are about 17.2\,m by 0.86\,m.)

Bonds between adjacent nylon panels were made using a glue recipe from Nyltech
of Acros Resorcinol and sodium meta bisulphite solution in water and ethanol.
Glue is sprayed as a fine mist onto both surfaces of the joint. The glue
chemically attacks a thin layer of nylon, making it tacky after a short
exposure time. The two panels are clamped under pressure in order to produce a
good seal and strengthen the bond.  Various combinations of exposure time, glue
amount, size of the droplets in the glue mist, applied pressure, and clamp time
were tested.  The optimal combination was found to be an exposure time of
several tens of seconds, pressure in excess of $3\times10^5$ Pa, and clamp time
greater than 4 hours.  Joints created with these parameters were mechanically
stronger than the nylon film itself.

After continued exposure to light, the optical properties of these joints
degraded slightly as the bond gained a greenish tint.  However, no
corresponding mechanical deterioration was observed.

\begin{figure}[h!]
\begin{center}
\epsfig{file=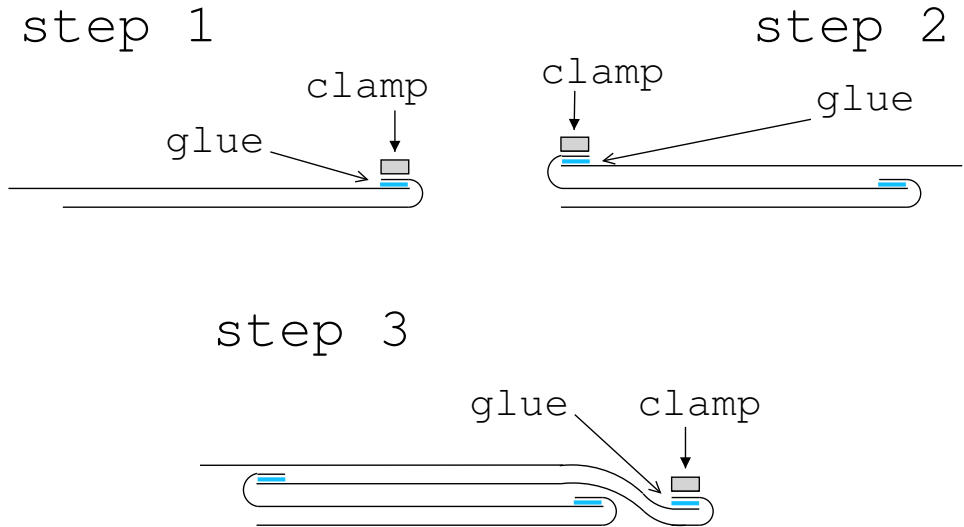, width=5in}    
\caption{The accordion-style stack technique used for constructing the nylon
vessel envelopes. The edges of wedge-shaped nylon panels are bonded together
to form the spherical vessel; the bond is temporarily clamped down against
the working surface to cure.  Panels are folded into a stack once bonded.
The process continues until all panels of a vessel are in
the stack. (In addition to permitting vessel construction within the limited
confines of the clean room, as a bonus the stacking permits the film to be
self-covering, reducing exposure to radon daughters in the air.)  The last glue
joint is conceptually identical but requires unfolding some panels in the stack
to align the top and bottom panels for gluing.}
\label{f:joint-sketch}
\end{center}
\end{figure}

\begin{figure}[h!]
\begin{center}    
\epsfig{file=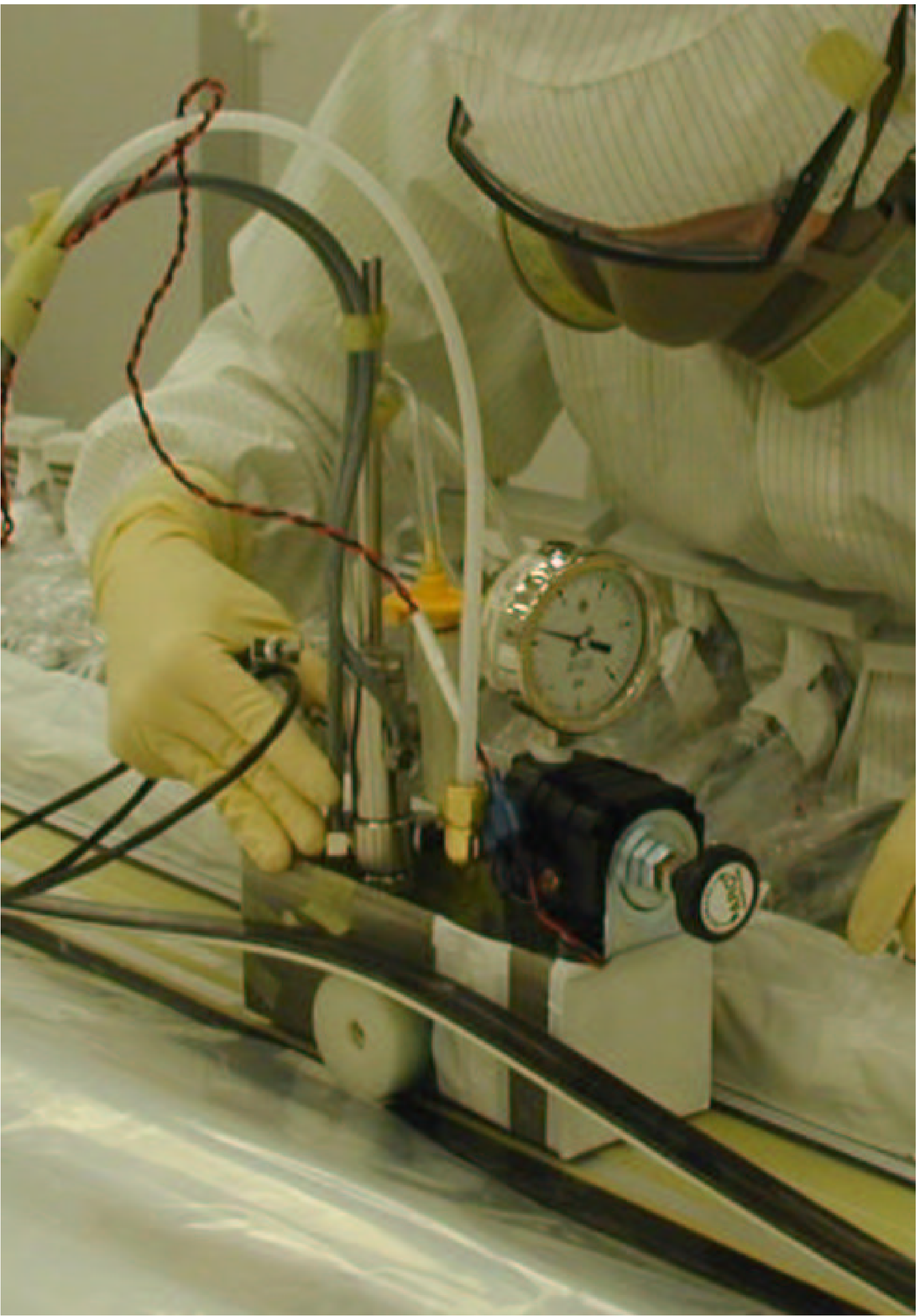, width=2.5in}
\hspace{0.2cm}
\epsfig{file=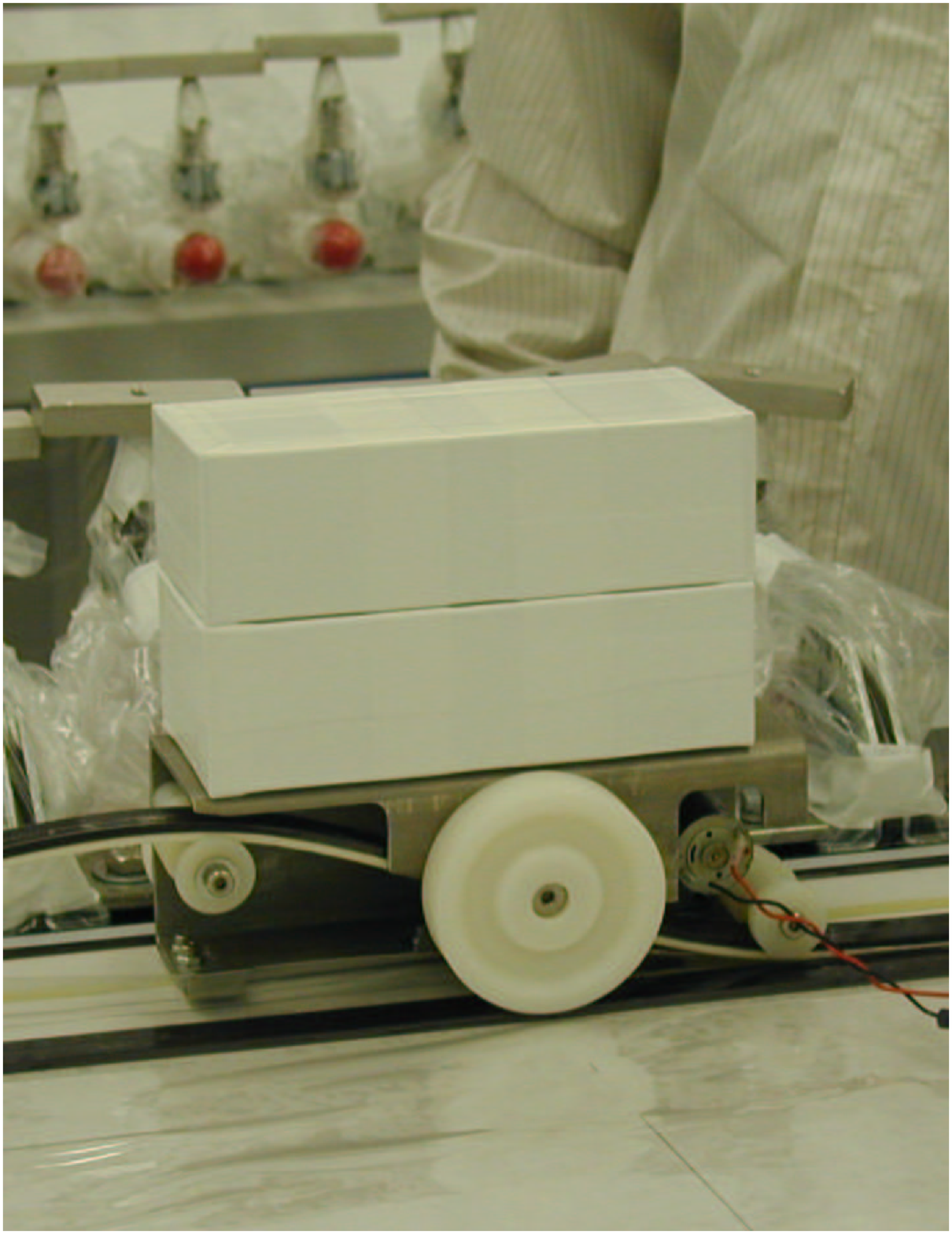, width=2.757in}
\caption{Making a glue joint for the nylon panels. A first cart (left) carries
the spray gun with which the resorcinol solution is applied and a second cart
(right) follows, folding the edge of the panel beneath onto the one above. The
clamps are then closed to apply the necessary pressure. Close-ups of the spray
and folding carts are shown.}
\label{f:carts}
\end{center}
\end{figure}

Adjacent panels were sequentially bonded as described above and folded over to
form an accordion-like stack (see Figure~\ref{f:joint-sketch}).  Care was taken
not to crease the nylon film at the folds.  While the glue set, pressure was
applied along the whole length of the seam with spring-loaded clamps mounted on
the assembly tables.  Remote-controlled carts were used to perform the cutting,
glue-spraying, and folding steps (see Figure~\ref{f:carts}).

Once all panels of the IV were on the stack, the last seam was made
by bonding together the edges of the first and last panels.  The polar end
regions were then put in place as described in section~\ref{ss:end-regions}.
The rope system was attached (section~\ref{ss:ropes}), followed by the optical
fibers, temperature sensors, and load cells (section~\ref{ss:instrumentation}).
The delicate operation of nesting the IV inside the OV was
then performed. One meter of the final OV seam close to each pole was bonded
first and the OV end regions were installed. After the IV was placed inside the
OV (within one of the top folds of the OV stack to be precise), the rest of the
OV seam was sealed. More details on the production operations are found
in~\cite{pocar}.

\subsection{The polar region design and tube assemblies}
\label{ss:end-regions}

The main challenge in building the polar end regions of the vessels consisted of
making the transition between nylon film and bulk material with a low-mass,
leak-tight, mechanically strong structure.  The low mass is necessary to reduce
the amount of radioactivity from material in the end regions.  At each pole of
the vessels is a set of tubes, the tube assemblies, connecting the vessels
to the SSS, water tank, and fluid handling plants. These
tubes carry the scintillator and buffer fluids into the IV and OV.
They also transfer the buoyant load from the nylon vessels to the
SSS.  The transitions between the nylon vessels
and the tube assemblies are by far the most delicate components of the system.

Three basic principles drove the design of the end regions:
\begin{itemize}
\item The end regions need to be mechanically stronger than the nylon membranes to which they are connected.  In the (unlikely) event of a catastrophic mechanical failure, we wanted to assure that the film envelopes would fail first for safety reasons, preserving the mechanical integrity of the supporting structure.
\item The end regions need to be leak tight; that is, the leak rate of the
end regions must not contribute a significant fraction of the overall leak
rate from the entire vessel.
\item The $\gamma$ ray-induced background resulting from radioimpurities
contained in the end regions should not contribute more than 0.1~events/day in
the 250--800\,keV energy range and for events localized in the innermost
100\,tons fiducial mass.  The corresponding requirements in terms of allowable
radioimpurities in the material for the IV end regions are extremely stringent.
Materials in the OV end regions have much less stringent requirements, due to
their greater
distances from the fiducial mass.  Great effort has been put into minimizing the
mass of the IV end region with the use of nylon film in place of bulk nylon and
in carefully selecting the construction materials.  Results are summarized in
section~\ref{s:radiopurity}.
\end{itemize}

\begin{figure}[h!]
\begin{center}
\epsfig{file=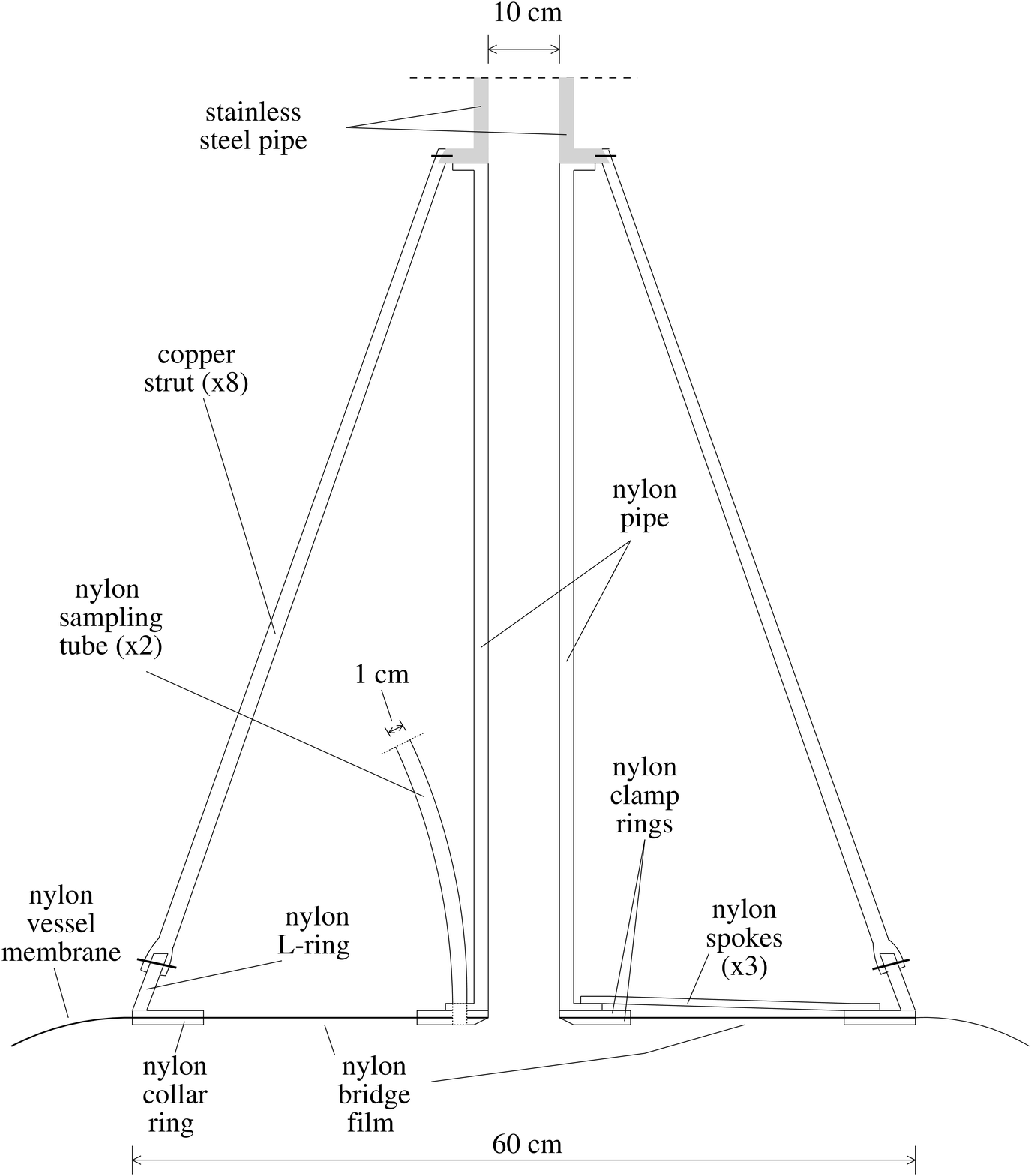, width=5in}
\caption{Sketch of the IV upper end region ({\it i.~e.}, polar) assembly in
cross-section. The gluing technique for transitioning from nylon film to bulk
nylon is shown in Figure~\ref{f:er-gluing-sketch}.}
\label{f:er-sketch}
\end{center}
\end{figure}

The IV end region is shown in Figure~\ref{f:er-sketch}. It consists of a nylon
ring, a circular nylon film cap, the central nylon tube that carries IV fluid,
and two flexible nylon tubes for direct fluid sampling and pressure
measurement. Six diagonal copper struts between the ring and the central tube
give mechanical strength against rotational and lateral stresses. Copper was
chosen for struts in the IV for its low radioactivity content and greater
strength than nylon.  The OV caps have a solid nylon disk instead of the nylon
bridge film and 8 stainless steel diagonal struts are used instead of copper
ones. The more complex design of the IV end-cap minimizes its radioactivity at
the smaller IV radius (the bridge film has significantly less mass than a bulk
nylon plate and is intrinsically less radioactive) and prevents obstruction of
light for events close to the IV poles. For the OV, stainless steel sampling
tubes are used instead of the more delicate nylon ones. 

The vessels and support structures are designed to withstand a load of 3000\,N.
This corresponds to a 5$^\circ$C temperature difference, or a 0.5\% density
difference,  between the scintillator and the buffer region.  The resulting
localized stress on the nylon membrane is about 40\,MPa (corresponding to a
vertical force of 300\,kg).  This assumes that the ropes capture most of the
1400\,kg buoyancy force and the film does not stretch.   However, the film will
stretch to relieve localized stress.   A finite element analysis shows that
after a short time the membrane stress will relax to $\sim$ 4\,MPa.  Some basic
findings are presented in section~\ref{ss:stress} below.\footnote{A more
detailed paper on the calculated stress levels of the nylon vessels is in
preparation.}

The assembly proceeded as follows: the IV nylon membrane, trimmed at the poles,
is glued with Resorcinol onto a 60-cm diameter bulk nylon ring that acts as a
collar on the inside of the vessel.  (In order to maintain good levels of
radiopurity, the bulk nylon pieces in the end regions were extruded, not cast.)
An annulus of transparent nylon bridge film, 0.5\,mm thick C38F nylon (the same
type used in the prototype Counting Test Facility vessel), connects the collar
ring to the nylon tube.  Note that C38F has mechanical properties similar to
those of Sniamid.  It is first glued onto the central nylon tube assembly
with the two flexible nylon sampling tubes and leak checked separately; it is
then glued onto the collar ring. Finally, a second 60-cm diameter nylon ring is
bolted onto the collar ring, clamping both the nylon vessel and the bridge film
membranes between them.  This second ring has a protruding lip, making it
L-shaped in cross section (an ``L-ring''), in order to provide anchoring points
for the copper struts, and to fix the midpoint positions for the ropes that
hold the vessels against buoyant and gravitational forces.

\begin{figure}[h!]
\begin{center}
\epsfig{file=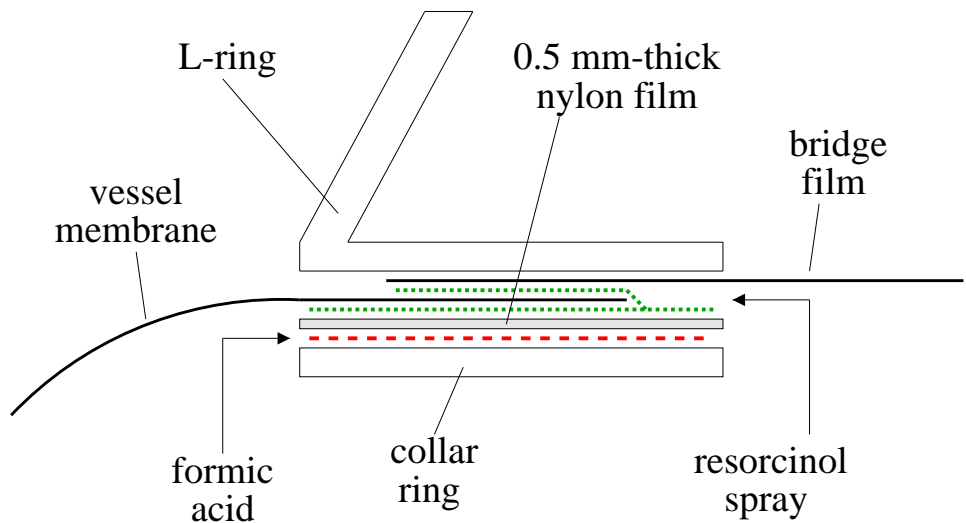, width=5in}
\caption{Gluing scheme for the end region transition between the vessel nylon
membrane and the polar assemblies. A layer of half millimeter-thick nylon film
is placed on the bulk nylon surfaces with pure formic acid (dashed line).  The
vessel membrane can then be glued to the assembly relying only on well-tested
resorcinol bonding (dotted line). The drawing refers to the IV collar ring
region, but this technique is applied to all the film-to-bulk nylon
transitions. The L-ring is simply bolted in placed (nylon bolts not shown for
simplicity).}
\label{f:er-gluing-sketch}
\end{center}
\end{figure}

The gluing technique is shown in detail in Figure~\ref{f:er-gluing-sketch}.  A
layer of C38F nylon film is glued onto the bulk nylon
surface using pure formic acid.  The bulk nylon surface is pre-treated with pure
formic acid to improve the bond between film and bulk.  Pressure is applied to
form a seal.  Thin nylon film is then Resorcinol-bonded onto the thick film as
described in section~\ref{ss:fabrication}.  After the vessel membrane is glued
onto the collar ring, the surface of the glue joint is cured with small amounts
of raw formic acid to smooth the step at each joint. Such features, if not
eliminated, would be natural channels for leaks.

The nylon tube and bridge film assembly were bonded similarly.  The nylon tube
section of the assembly had its surfaces treated with formic acid, then glued
together with nylon and formic acid paste. The two 3/8-inch flexible nylon
lines at each pole used for fluid sampling and pressure measurement were then
added, attached by gluing in place with formic acid and a nylon Swagelok part.
Each end tube was leak checked during assembly by pressurizing the system,
mounted on a frame made for the purpose, with SF$_6$ and sniffing for it from
the outside.  Mechanical and leak checking tests of prototype end region
assemblies, as well as the final pieces, under the tensile design stress of
3000\,N, exceeded the required specifications by many orders of magnitude.

The OV end regions are similar to the IV ones, but simpler.  They do not
incorporate a bridge film, since they obstruct much less scintillation light
than the IV end regions would block if opaque.  Instead, an L-ring is attached
to a bulk nylon plate, which in turn connects to the stainless steel section
of the central tube.  The connection is made with an O-ringed flange and is
performed during the vessel nesting step.  No leak checking test was carried
out for the actual OV end regions.

The two sets of tube assemblies are designed to transport fluids separately to
each of the three volumes defined by the two vessels and the SSS, and in
addition to keep the nylon vessels suspended in the center of the SSS, rigidly
fixing their positions with respect to it.  The tube assemblies must take the
load of the vessels, while having stringent radiopurity requirements due to
their proximity to the fiducial volume. In addition they accommodate fluid
sampling lines and allow the passage of instrumentation lines for monitoring
the vessels.

Connection of the tube assemblies to the SSS was implemented with double viton
O-ring gaskets. A rotatable flange couples the south tube assembly and the
south 1-meter flange of the SSS, in order to accommodate differing orientations
of the north and the south 1-meter flanges.

\begin{table}[t]
\caption{\label{t:EndRegionRad}Measurements or upper limits of radioactive
contaminants in the end region
components that contribute most to the background.  Radioactive contaminant
levels are reported in parts per trillion (ppt) by mass for U and Th,
parts per million (ppm) by mass for potassium (all isotopes), and as
activities in mBq per kg for $^{60}$Co.  As most of these measurements
were performed by early 2001, and the half-life of $^{60}$Co is 5.3~years, we
expect that the current $^{60}$Co contamination levels are reduced by
more than a factor of two.}
\begin{center}
\small 
\begin{tabular}{lrrrr}
\hline\hline
Component & $^{238}$U & $^{232}$Th &  K$_{nat}$ & $^{60}$Co \\
          & [ppt]     & [ppt]      &  [ppm]     & [mBq/kg]  \\
\hline
Nylon rings (extruded)   &  $<$ 30 &  $<$ 50  & 0.13 & $\ll$ 1 \\
Nylon tubes (cast)       &  $<$ 50 & $<$ 50 & $<$ 0.70 & $<$ 2 \\
Copper struts            & $<$ 30 & $<$ 50 & $<$ 0.02 & $\ll$ 1 \\
Steel flanges            &  150 & 920  & $<$ 0.15 & 15 \\
Steel tubes              &  150 & 700  & $<$ 0.26 & 10 \\
Steel OV mount rings     &  140 & 2100 & $<$ 0.13 & 20 \\
Steel OV flanges         &  160 & 3300 & $<$ 0.07 &  2  \\
\hline
\end{tabular}
\end{center}
\end{table}

Table~\ref{t:EndRegionRad} summarizes the radioactive contamination of
materials in the end regions~\cite{lowrad}.
We found that extruded nylon has in general a lower
K content than cast nylon, thus opted for extruded nylon in the components
closest to the fiducial volume.  An unexpected surprise was the level of the
isotope $^{60}$Co in the steel parts.  This anthropogenic
isotope, which is often used for profile radiography in
monitoring stainless steel production, decays by
$\beta$ emission with the production of 1.17 and 1.33\,MeV $\gamma$ rays.
Since the half-life of $^{60}$Co is 5.3~years, it is presumed that the levels
actually present in Borexino once the detector is fully operational will be
lower than those reported here.

Electropolished stainless steel is used for the outermost parts of the tube
assemblies where they attach to the SSS; low activity nylon is employed
between the two vessels.  Due to the rigid radiopurity requirements, the steel
portions of the tube assemblies have been welded with thorium-free rods.  All
the tube assemblies were pickled, passivated and precision cleaned before
mounting.

\subsection{The ropes}
\label{ss:ropes}

Each vessel is restrained
by two sets of ropes to prevent vertical motion in each direction.  Each set
consists of 18 ropes for the IV and 20 ropes for the OV.
The ropes wrap around the vessel longitudinally, with the
rope ends attached to the stainless steel part of the tube assemblies.
The center of each rope is routed through the L-ring opposite the rope ends.
Nominal rope lengths are 27.526\,m for the IV and 35.072\,m for the
OV. The ropes were pre-stretched, ensuring they were completely taut,
before being cut to these lengths.
For monitoring the tension in the ropes, load cells are employed
where the ropes attach to the fluid tube.  The tension in each IV
rope would be 59\,kg with a density difference of 0.5\%.

In case the panels of the vessels were not 
aligned perfectly longitudinally, it was decided to allow the
ropes to float freely over the vessel surfaces rather than being fixed to
individual panels.   This prevents the ropes
from contributing a sideways component of force to the panels.  
In order to prevent several ropes from bunching up along one meridian of a
vessel, all of them are attached to a perpendicular set of ropes.
These horizontal ropes (seven on the IV and nine on the OV) are fixed to the
vessels with tabs on the nylon panels.  Thus the complete set of ropes forms a
grid-like net over the vessels.

The choice of material for all of these ropes is critical due to stringent
requirements for mechanical strength and low levels of radioactive
contaminants.  The rope used for Borexino is made of multistrands of Tensylon
with an ultimate yield strength of 590\,kg.  Tensylon is an ultra-high
molecular weight polyethylene manufactured by Synthetic Textiles.  It
has a tensile strength of 1.45$\times 10^9$\,Pa, and a Young's modulus
of 9.6$\times 10^{10}$\,Pa.  The
diameter of the ropes was chosen such that the maximum load would be 10\%
of the yield strength.  Lifetime under this continuous
stress is projected to be more than 10 years, but should be much longer under
expected loads.  Failure would eventually occur due to creep, an irreversible
slow change in the arrangement of polymer chains under load.  This creep
has been measured for Tensylon to be unaffected by soaking in pseudocumene.

Potassium salts are typically used in the coating of nylon ropes, but must be
avoided due to the presence of the naturally occurring
radioactive isotope $^{40}$K.  Tensylon rope (woven by Cortland
Cable from fibers produced by Synthetic Textiles) is uncoated.
Its total potassium content (all isotopes) was measured to be on the order of
0.2\,ppm.  Heavy metal contaminants in the Tensylon are also lower than or
comparable to other ropes; see Table~\ref{t:rope-contamination}.

\begin{table}[t]
\caption{\label{t:rope-contamination}Radioactive contaminants of various
ropes.  The rope selected for Borexino was the Tensylon.}
\begin{center}
\small
\begin{tabular}{lrrr}
\hline\hline
Sample & $^{238}$U & $^{232}$Th & K$_{nat}$ \\
       & [ppt] & [ppt] & [ppm] \\ \hline         
Vectran, 1200 lbs.         & $<$ 100 & $<$ 30 & 45.0   \\
Zylon 13                   & $<$ 50 & $<$ 50 & 10.2 \\
Spectra, 300 lbs.          & 60     & 500     & 7.5  \\
Aramid Technora T-000      & \multicolumn{2}{c}{\it not measured} & 5.7  \\
Berkeley nylon monofilament& $<$ 90  & $<$ 150  & $<$ 0.2  \\
Tensylon by Cortland Cable & $<$ 70 & $<$ 160 & 0.5 \\
Tensylon (final batch)     & $<$ 95 & $<$ 180 & 0.2 \\
\hline
\end{tabular}
\end{center}
\end{table}

\subsection{Instrumentation}
\label{ss:instrumentation}
A critical part of the vessel assemblies, although not directly related
to containment of the scintillator, is the instrumentation.  This encompasses
a variety of sensors and calibration devices that measure temperature,
rope strain, vessel position, and other local conditions.

\subsubsection{Temperature sensors}
\label{sss:temperature-sensors}

The maximum temperature difference across the nylon vessels must be less than 
$\Delta T = \pm$ 5$^\circ$C.  The density difference due to
temperature swings can account for most of the buoyant
force, as $\Delta\rho/\Delta T \approx 8 \times 10^{-4}$\,ton/m$^3$/$^\circ$C
for both the pseudocumene and PPO scintillator, and the pseudocumene and DMP
buffers.  In order to
check for temperature gradients, a set of temperature monitoring sensors 
are installed.

On the vessels, four temperature sensors are mounted at each end, north and
south.  Two measurement points at each end are in the outer buffer volume: one
set at the level of the SSS, and one halfway between the SSS and the OV.  Two
sensors are mounted at each end in the inner buffer volume, one at the level
of the OV and a second 35\,cm closer to the IV.  These
positions permit the measurement of vertical temperature gradients.

The temperature sensors chosen for the nylon vessels are custom made Pt100 RTDs
produced by Fisher Rosemount, Italy.  The sensors are Pt resistors, nominally
100\,$\Omega$ at a temperature of 20$^\circ$C.  Their sensitivity is rated at
$<\, 0.1^\circ$C, better than the requirement of $\pm$0.2$^\circ$C.  To read
the sensors accurately, four copper wires are used to cancel the resistances of
the cables themselves.  These readout wires are embedded in MgO insulation;
they are surrounded by a welded stainless steel conduit that carries them
to the exterior of the detector.  The mass per unit length of the conduit is
40\,g/m.

\subsubsection{Load cells}
\label{sss:load-cells}

Load cells were installed to measure the buoyant force on the vessels due to
differences in temperature and composition between the three fluid volumes.  At
a constant temperature, the buoyant upward force on the IV due to the
difference in densities between scintillator and buffer fluids is
$\sim$280\,kg~\cite{pocar}.  With a $\pm5^\circ$C temperature difference across
the IV, the buoyant force on the IV goes from 1.58~tons in the upward direction
to 1.02~tons in the downward direction. With the same temperature difference
across the OV, there is a buoyant force of 2.45~tons.  (The weight of the nylon
vessels, strings and other mechanical parts is negligible.)

Buoyant forces on the vessels are transferred to the ropes and
the tube assemblies.  The force on both sets of components is
measured using load cells.   The selected load cell is the Sensotec model 34,
which has a mass of 70\,g.
Each cell has a hysteresis no greater than 0.2\% of the full-scale, and
non-repeatability no greater than 0.05\%.  The temperature dependence of
the cells is 0.005\% of full scale per $^\circ$C.  Full scale (which may
be tuned) is set at 500\,kg force for the IV load cells, and 1000\,kg force for
the OV load cells.

The hold-down ropes are grouped together at their ends 
and attached to load cells.  At the north and south pole of the
IV, each of the 6 groups of 6 rope ends is connected to a load
cell attached to the tube assemblies (for the OV, there are 8 groups of 5
rope ends at each pole).  When the vessels are fully inflated,
the load cells are attached to the central tube at angles of 58$^\circ$ and
53$^\circ$ from the vertical for the IV and OV, respectively.
The maximum force expected on each load cell is thus 495\,kg (510\,kg) for the
IV (OV).

The maximum force expected on the tube assemblies that support the vessels was
calculated with a Finite Element Analysis simulation to be about 240\,kg
force. 
Load cells similar to those on the ropes, but with a lower capacity,
are attached to the tube assemblies to measure this strain.
There are two such load cells for each of the four tubes.

\subsubsection{Calibration light sources and optical fibers}

Since the two nylon vessels are not rigid, observing their precise shapes and
positions may be important when defining fiducial volume cuts or making Monte
Carlo simulations of external radiation.  This monitoring ability is also one
means by which the safety of the vessels can be ensured.  In order to
accomplish this, CCD cameras will be used to observe point-like light sources
mounted on the outside surfaces of the vessels. The sources are 3/32-inch
diameter teflon diffuser spheres attached to fiber optic cables, which are
connected to a laser outside the detector.  There are 38 light diffusers on the
IV and 40 diffusers on the OV.

Eight additional optical fibers mounted on the IV are used for the purpose of
monitoring scintillator quality. They use 355\,nm light to excite the PPO in
the scintillator near the membrane.  A full description of the light diffusers
and optical fibers, including their precise positions on the vessels' surfaces,
is provided in chapter~7 of reference~\cite{pocar}.

\subsubsection{Feed-through connections}
\label{sss:break-out}

To prevent leaks of the pseudocumene,
the holding system is designed so that all fluids
are doubly contained.   Electrical cables from the instruments, optical
fibers to the diffusers, and fluid sampling tubes from the vessels must
therefore be routed through a complicated double set of feed-throughs.

The 44 electrical signals from temperature sensors and load cells run
through wires completely contained in bellows.  Wires run from these bellows
through a conflat flange to meet a hermetically sealed feed-through
isolated from the pseudocumene.
There is also an individual feed-through for each of the 86 optical fibers.

To maintain a good seal where tubing from the vessels leaves the SSS,
single tubes are welded directly to the flange coupling, which
are joined with the feed-through fittings outside the SSS with Swagelok
connectors.  The connections of the tubes to the nylon vessels were sealed with
metal gaskets.  Seals between the tube assemblies and SSS were made with double
viton o-ring gaskets for each tube.

\subsubsection{Radioactivity measurements for instrumentation elements}

It was necessary to ensure that radioactive background from the temperature
sensors, load cells, and optical fibers placed inside the detector was
sufficiently low.  Radiopurity measurements of these components are detailed in
Table~\ref{t:instrument-cont}.  The primary contribution from the temperature
sensors is due to the long stainless steel conduits that go out through the
feed-throughs.  Most of the lengths of these conduits are far from the IV;
hence the relatively high levels of radioactive contaminants are not as
damaging as one might think at first look.  It should also be noted that the
measured load cell was not equipped with a Swagelok
connector.  However, the stainless steel Swagelok is not expected to make a
large contribution.

\begin{table}[t]
\caption{\label{t:instrument-cont}Radioactive
contaminants in the instrumentation.  The tabulated mass of 1.1\,kg for the
load cells refers to only the 16 that are positioned near the IV.}
\begin{center}
\begin{tabular}{lrrrcr}
\hline\hline
Component & Mass & $^{238}$U & $^{232}$Th & K$_{nat}$ & $^{60}$Co \\
 & [g] & [ppt] & [ppt] & [ppm] & [mBq/kg] \\ \hline
\multicolumn{5}{l}{\it Light diffusers and optical fibers} \\
Teflon beads		& 1	& 18	& 29	& 13 & - \\
Resorcinol		& $<$ 1	& 26	& 100	& 0.051 & - \\
Formic acid		& $<$ 1 & $<$ 2	& $<$ 2	& 0.014 & - \\
Optical fibers		& $\sim$500 & $<$ 50 & $<$ 30 & 0.135 & - \\ \hline
Load cells		& 1120	& $<$ 4500 & 2000 & $<$ 1 & 14 \\
Temperature sensors	& 40\,g/m & 33000 & 1900 & 3.4 & 9 \\
\hline
\end{tabular}
\end{center}
\end{table}

\subsection{Leak tightness specifications and leak checking}
\label{ss:leak-tightness}
The assumption of a 10~year experimental lifetime with no fluid purification
sets the leak tightness specification for the entire IV assembly.
The tolerable leak rate was determined by the amount of PPO allowable in the
buffer-quenched volume (inner buffer) between the two vessels.  DMP leaking into
the IV is much less worrisome.  PPO concentrations in the inner
buffer must be below 10\,ppm; otherwise the overall trigger
rate of the experiment would be too high~\cite{bx-quenching}.  The IV
therefore has a leak tightness specification of $< \, 10^{-2}$~cc/s of
pseudocumene 
at a pressure difference of 1\,mbar.  This is equivalent to 1\,cc/s of
gas at the same relative pressure.  (Strictly speaking, the equivalence is
true only for viscous, {\it i.~e.}, not molecular, leaks, as the scaling from
pseudocumene to gas is done using the ratio of viscosity coefficients.)

Since the leak rate in the OV is much less critical, the OV leak rate
specification was conservatively set at $< \, 1$\,cc/s/mbar of pseudocumene.
Leak specifications for the individual end regions were set to no more than
a fraction of the overall vessel specifications, specifically,
10$^{-3}$\,cc/s liquid leak rate for the IV end regions and 0.1\,cc/s for
the OV end regions at 1\,mbar overpressure.

The tightness of the bonding technique was tested by bonding together two 3\,m
long panels to form an inflatable, cigar-shaped balloon.
For testing, the prototypes were
filled with air and a trace amount of SF$_6$ gas.  Leaks were measured with a
residual gas analyzer sniffing near the joints.  The tests were done at 50\%
relative humidity, and the film was creased and crinkled severely before
inflation.  No noticeable leak increase was observed within a sensitivity of
$\sim 10^{-5}$\,cc/s of pseudocumene, scaled to the size and operational
differential over-pressure of the Borexino IV of 1\,mbar. To a good
approximation, gas and liquid leak rates differ by the ratio of the
viscosities. The liquid leak rates quoted above are thus approximately 100
times smaller than the explicitly measured gas leak rates.

\begin{figure}[h!]
\begin{center}
\epsfig{file=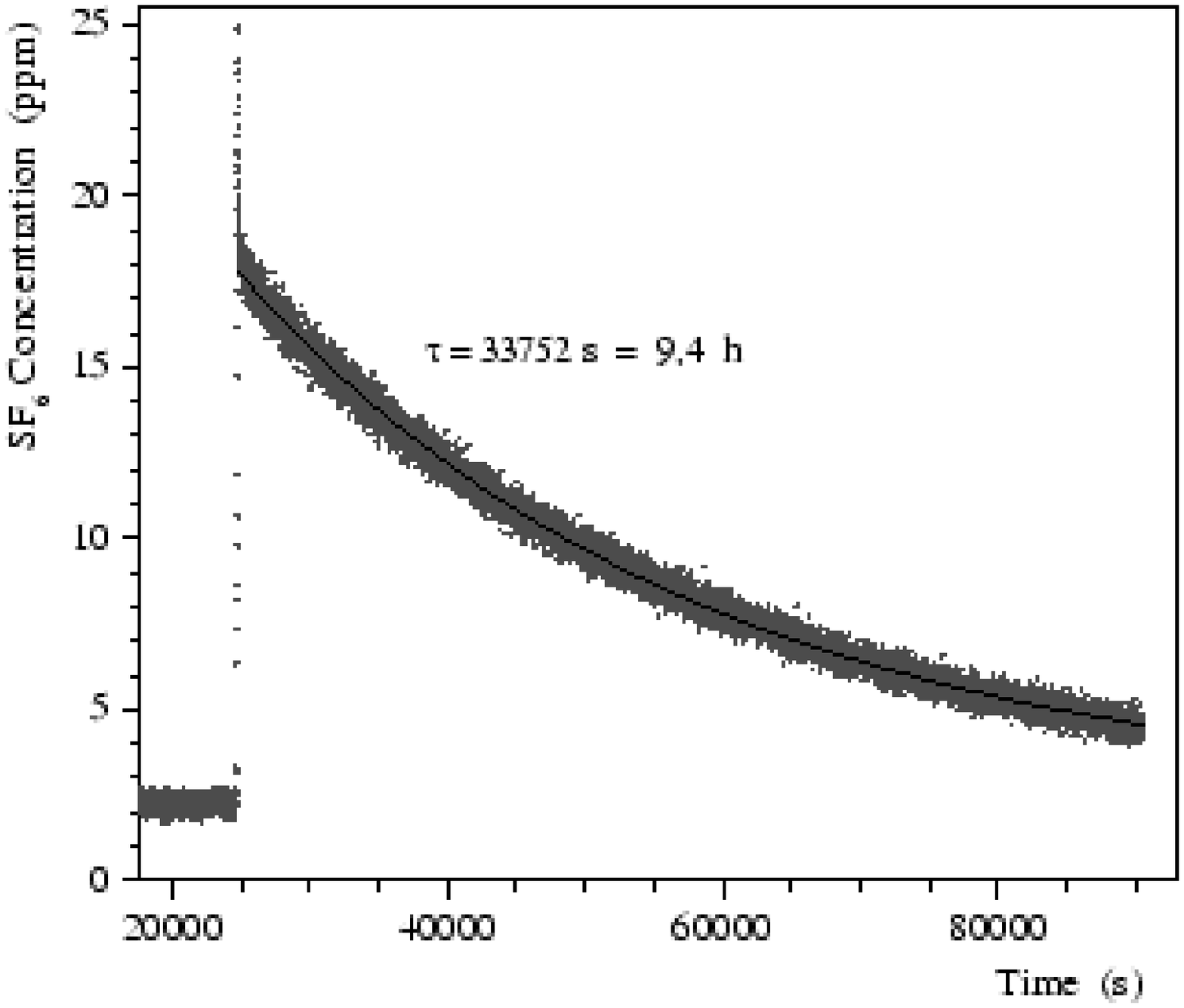, width=2.6in}
\epsfig{file=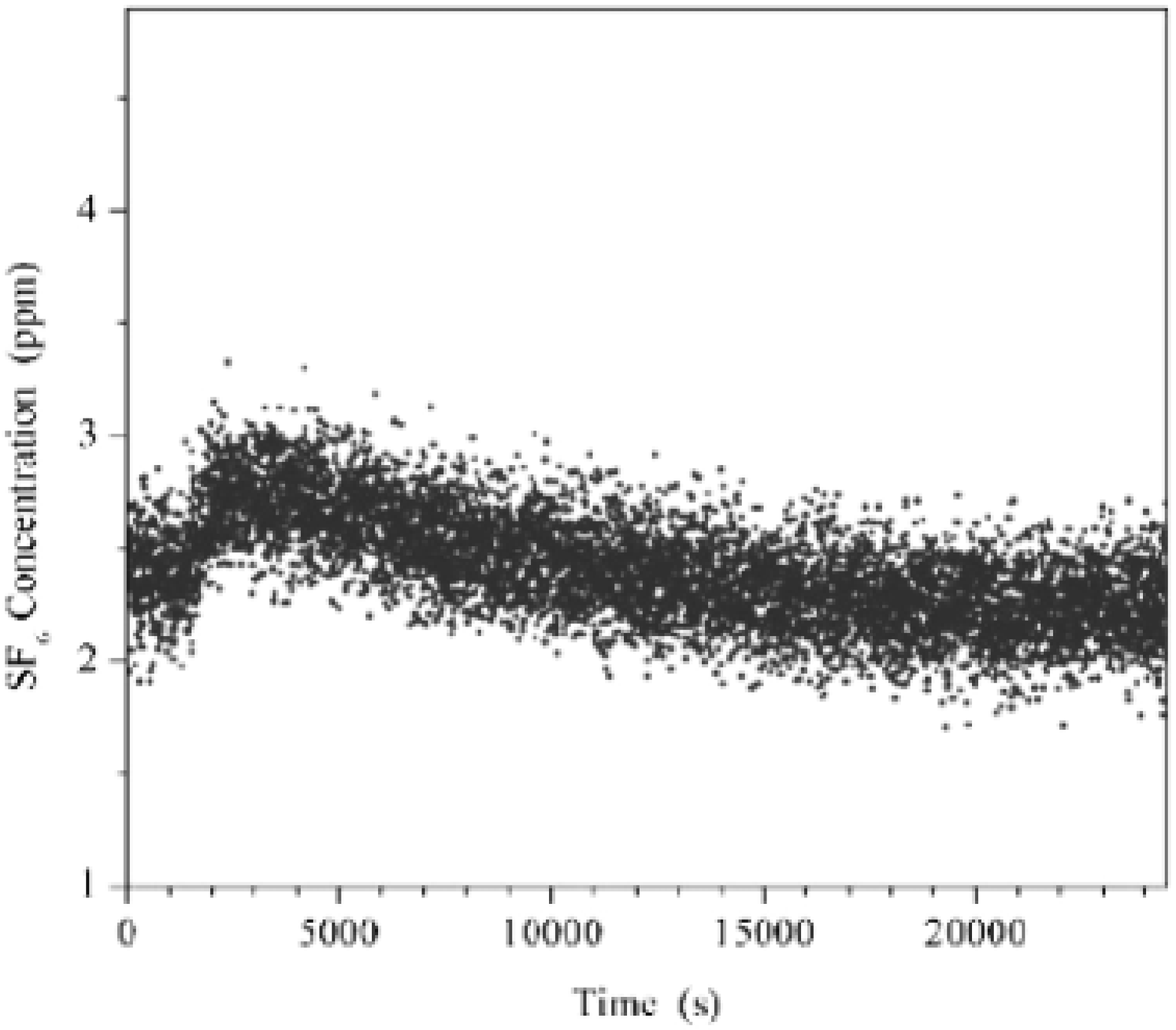,width=2.8in}
\caption{Leak checking the IV prior to shipping. Left: mixing of SF$_6$ gas
that was injected directly in the clean room, and the decay time of its
concentration, due to dilution with new air entering the clean room, following
an injected spike.  The plot shows that the mixing time in the clean room is
much faster than the dilution time.  Right: the step in SF$_6$ concentration
that occurs after an accumulation bag surrounding the SF$_6$-filled nylon
vessel is flushed into the clean room.  The size of this step after a 10 hour
accumulation in the bag implies an equivalent leak rate of
10$^{-3}$\,cc\,pseudocumene/s/mbar for the IV. }
\label{f:iv-lc-table}
\end{center}
\end{figure}

The entire IV assembly was also tested shortly after fabrication.  Once the IV
was completed with the addition of the endrings (section \ref{ss:end-regions}),
a table top leak test was performed. The folded IV was first slightly inflated
with humidified pure SF$_6$ gas to an over-pressure of about 0.1\,mbar. A thick
layer of polyethylene film was placed around the vessel stack and tightly taped
to the table, forming a closed volume in which a leak of SF$_6$ would
accumulate. The cover layer was then lifted and the concentration of SF$_6$ in
the sealed clean room measured as a function of time. Gas mixing in the clean
room is essentially instantaneous (Figure~\ref{f:iv-lc-table}, left), so the
SF$_6$ concentration step and the volume of the clean room give the amount of
gas that has leaked. Consideration of the accumulation time and the
over-pressure gives the leak rate. The measured leak rate was
10$^{-3}$\,cc/s/mbar of pseudocumene; this value was 10 times better than the
design goal (Figure~\ref{f:iv-lc-table}, right).

In addition, the vessels were also leak checked once they were installed
at Gran Sasso; see section~\ref{ss:leak-rate-tests} for details.

\subsection{Calculated stresses on the nylon film}
\label{ss:stress}
In the Borexino design, the nylon vessels sit in a nearly buoyancy-free environment.
The presence of different additives in the pseudocumene inside
(1.5\,g/liter PPO) and outside (5.0\,g/liter DMP) the IV induces a
density difference across the film, with scintillator 0.09\%
less dense than buffer fluid~\cite{pocar};
this gradient could be enhanced to a few tenths of a percent in the
unlikely scenario of a uniform and relatively large temperature difference between the fluids inside
and outside the vessel.  The resulting buoyant force
would pull the vessel up or down, depending on the direction
of the density gradient, and
reaction forces would develop at the contact with the
hold-down ropes and the junction of the film to the end regions. 
As a result, in-plane stresses pull on the nylon membrane.
The membrane responds elastically, up to a certain
point, and then yields to creep until equilibrium is reached.

In order to evaluate the membrane stress and whether it is a challenge to the
vessel structural integrity, we performed a static Finite Element Analysis (FEA)
using version 6.1 of ABAQUS/Standard, a commercial 
Finite Element Algorithm for non-linear structural analyses. 
Under the approximations of neglecting friction and creep,
the problem was also addressed analytically with the EMsolver code~\cite{fea}.
When the FEA model (which
also does not take frictional forces directly into account) was
tweaked to exclude the effects of creep, the results were in good agreement.

The structural analysis exploits the cyclic symmetry of the vessel around 
the $z$ axis and only models one nylon film panel, 
with glued seams at the edges and two ropes (section~\ref{ss:ropes}) running over its middle, one wrapped 
around the top and one around the bottom.
The analysis is performed in a spherical coordinate system, with in-plane
stresses
considered along the meridional ($\theta$) and hoop ($\phi$) directions. 
Symmetry around the $z$ axis is enforced by the ABAQUS Multi
Point Constraint, which equates 
radial, circumferential and axial displacement components of the nodes
on the left and right edges of the film panel. 
In addition, all translational and rotational degrees of freedom of
the end regions are constrained, while the boundary between end region and 
nylon membrane retains rotational degrees of freedom.
The ropes are not allowed transverse motion.

The analysis models the end regions with 1\,cm nylon shell elements, the ropes
with beam elements of 1.75\,mm radius and the mechanical properties of
Tensylon, and the internal volume of scintillator as an incompressible fluid of
889\,kg/m$^3$ density inside the vessel. 
The nylon envelope is modeled with
\emph{membrane elements}, thin surfaces that offer in-plane strength 
but no bending stiffness.
We used a realistic initial overpressure $P_0=50$\,Pa.
As the density gradient induces a buoyant load, which in the following
we assume to be upwards, 
the incompressible fluid reacts by increasing its 
own internal pressure, from the initial value $P_0$ to a final value $P_{1}$.
The resulting pressure differential across the membrane is a positive number  
(points outwards), depending on the z coordinate, which for
  $z=0$ at the bottom, $z=h=8.5\,$m at the top, is 
  $P(z) = P_{1} + \Delta\rho g z$ with $z \in [0,h]$.

For large membrane stress ($>10\%$ of the tensile yield), the 
elastic model and Hooke's law lose accuracy. 
The nylon film response to external loads is viscoelastic, 
as the molecules realign themselves and the stress-strain relation 
 changes in time, as described by the \emph{normalized creep compliance}:
\begin{equation}
j(t)= 
E_0\;\frac{\varepsilon(t)}{\sigma_0}
\end{equation}
where $\varepsilon(t)$ is the time-dependent
strain associated with a constant stress $\sigma_0$ 
and $E_0$ is the Young's modulus. 
In a linear viscoelastic material, the normalized creep compliance
is not a function of the applied
stress $\sigma_0$.
We modeled the film creep with the ABAQUS quasi-static, viscoelastic analysis 
option. We treated nylon as a linear viscoelastic material,
which is realistic up to about 30\% of the tensile yield,
with the creep compliance measured
under uniaxial tension at 20\% relative humidity.

It must be noted that wrinkling (caused by regions of compressive stress) and
creep cannot be simultaneously treated by our model.  A separate analysis that
takes wrinkling but not creep into account, using the uniaxial tension
mode, 
has also been performed, though there is insufficient space to discuss the
results in detail.  We found that in the worst-case
scenario, the stress in the film rapidly causes creep; hence creep cannot be
neglected and the analysis needs to be viscoelastic.  If the vessel starts out
slightly over-inflated (we assumed 50\,Pa), wrinkling becomes a very small
effect and the viscoelastic analysis without creep does a good job of
simulating the vessel.

\begin{figure}[h!]
\begin{tabular}{cc}
\epsfig{figure=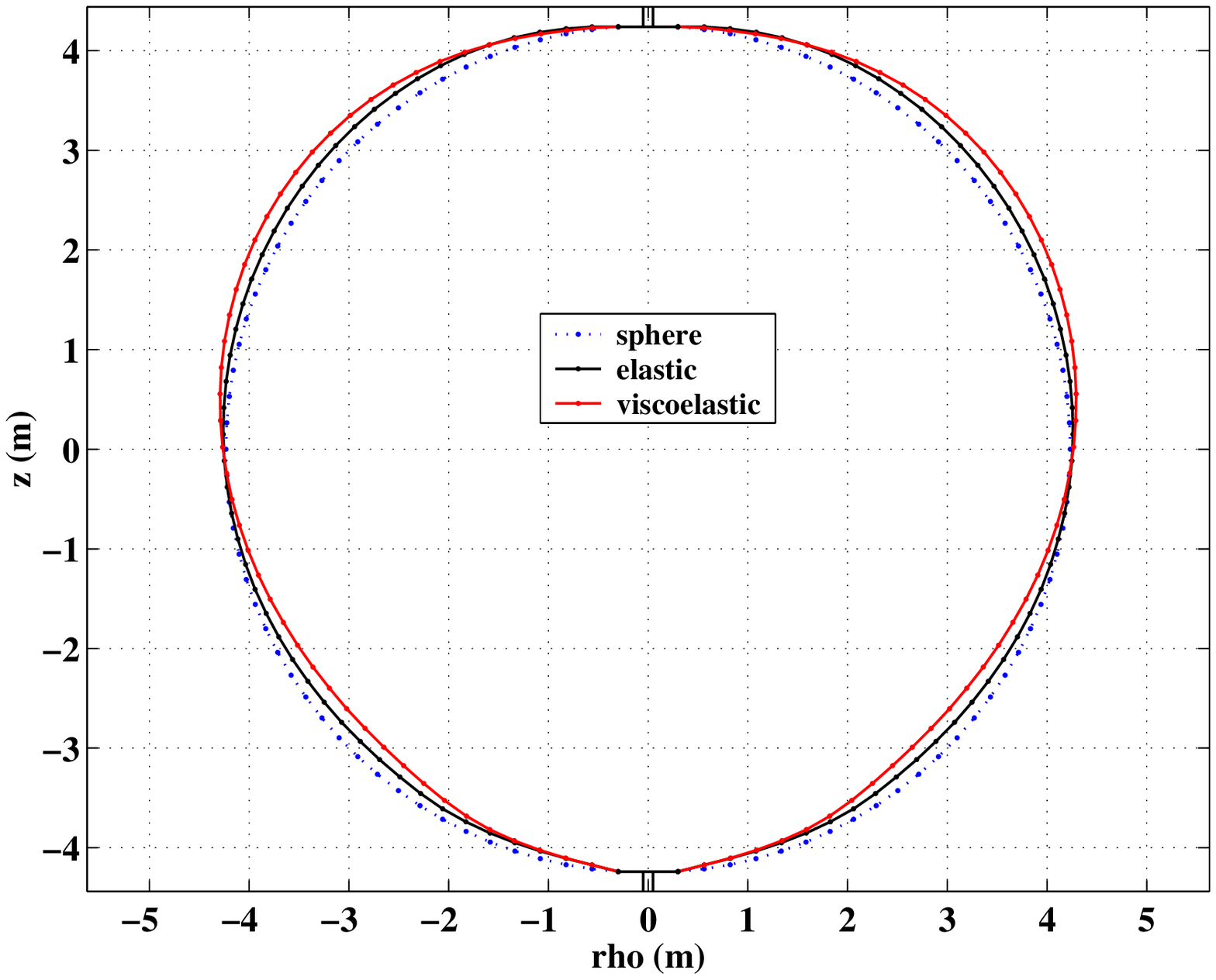,width=2.75in}&
\epsfig{figure=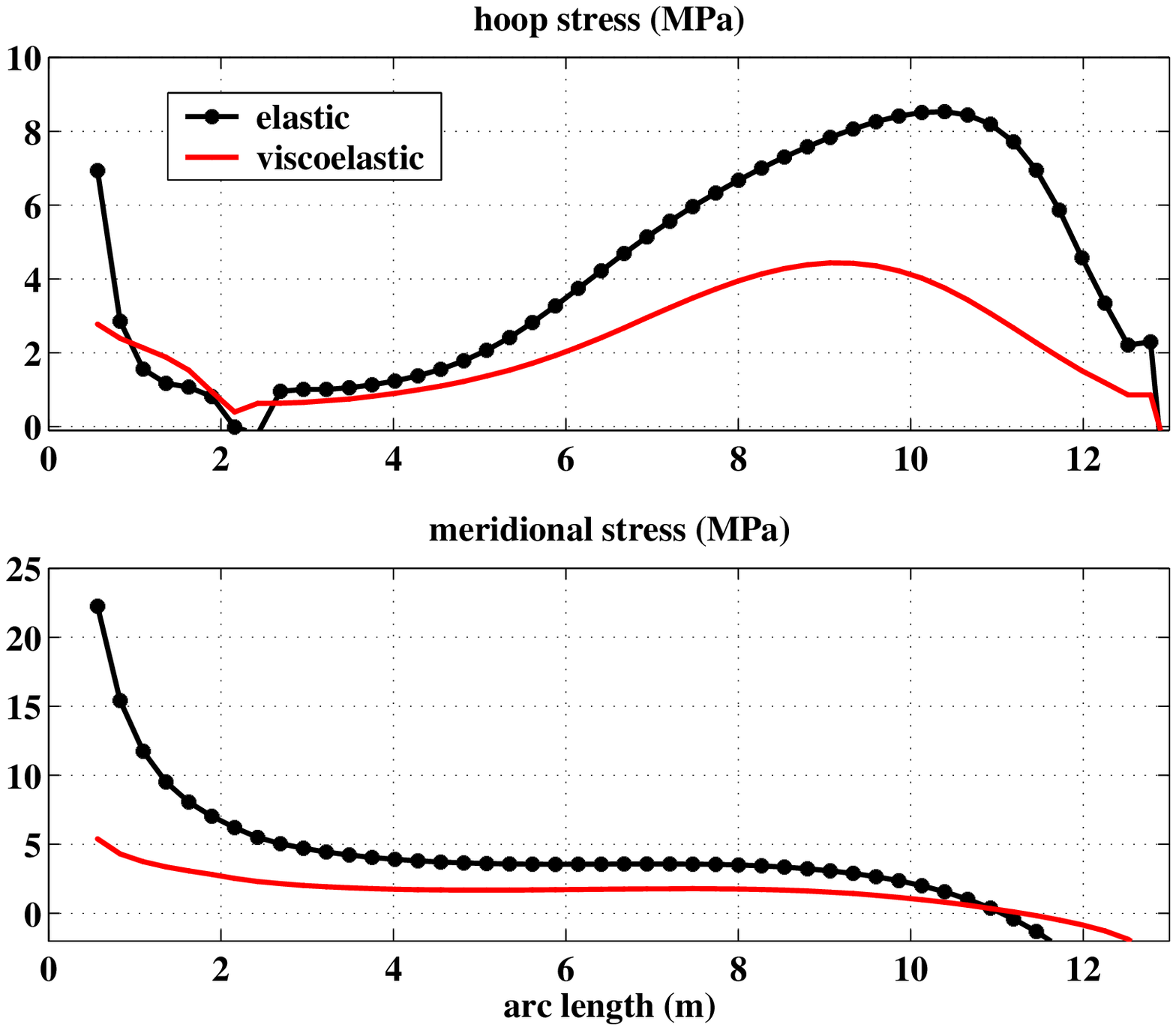,width=2.75in} \\
\multicolumn{2}{c}{ 
20\% relative humidity} \\
\epsfig{figure=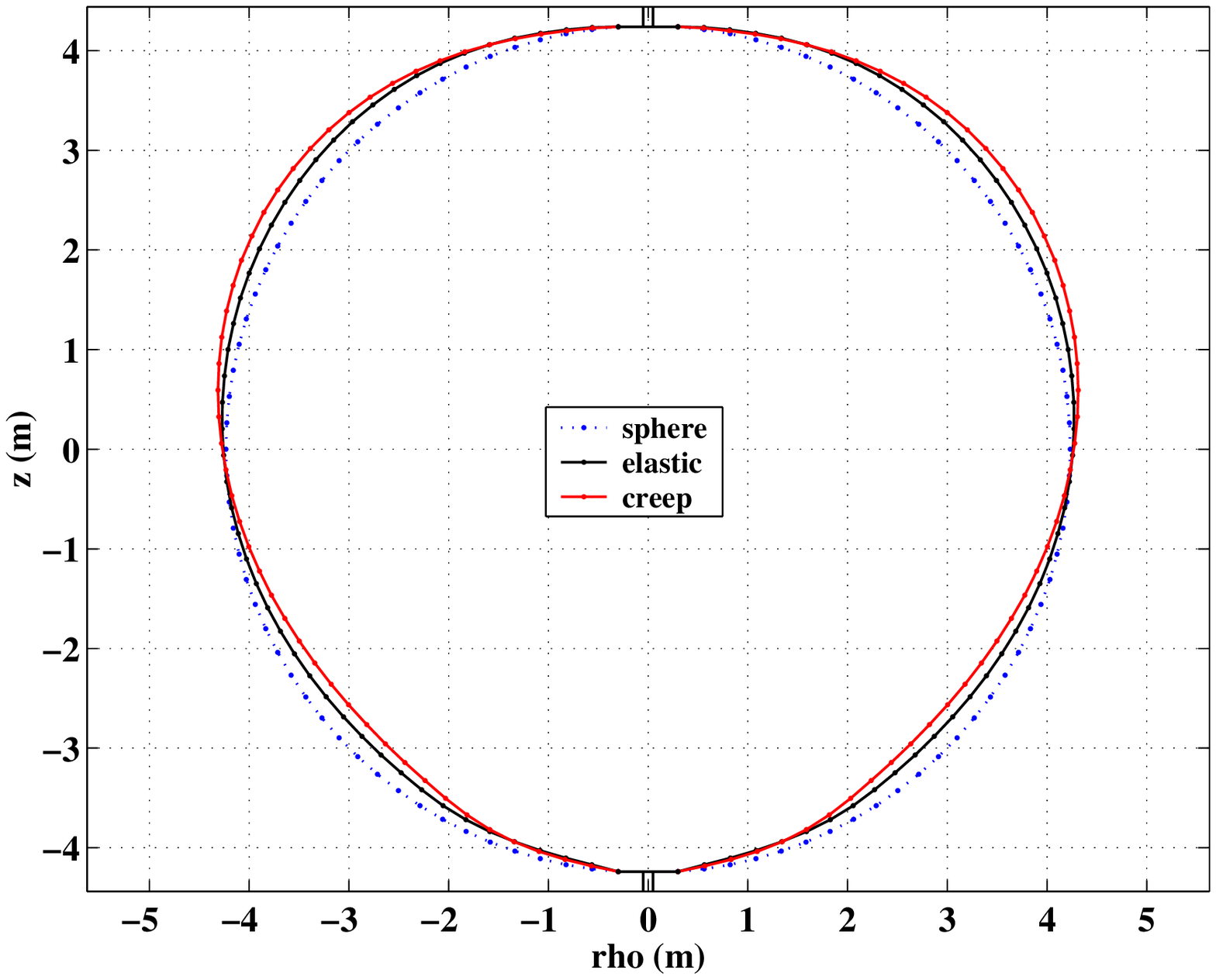,width=2.75in}&
\epsfig{figure=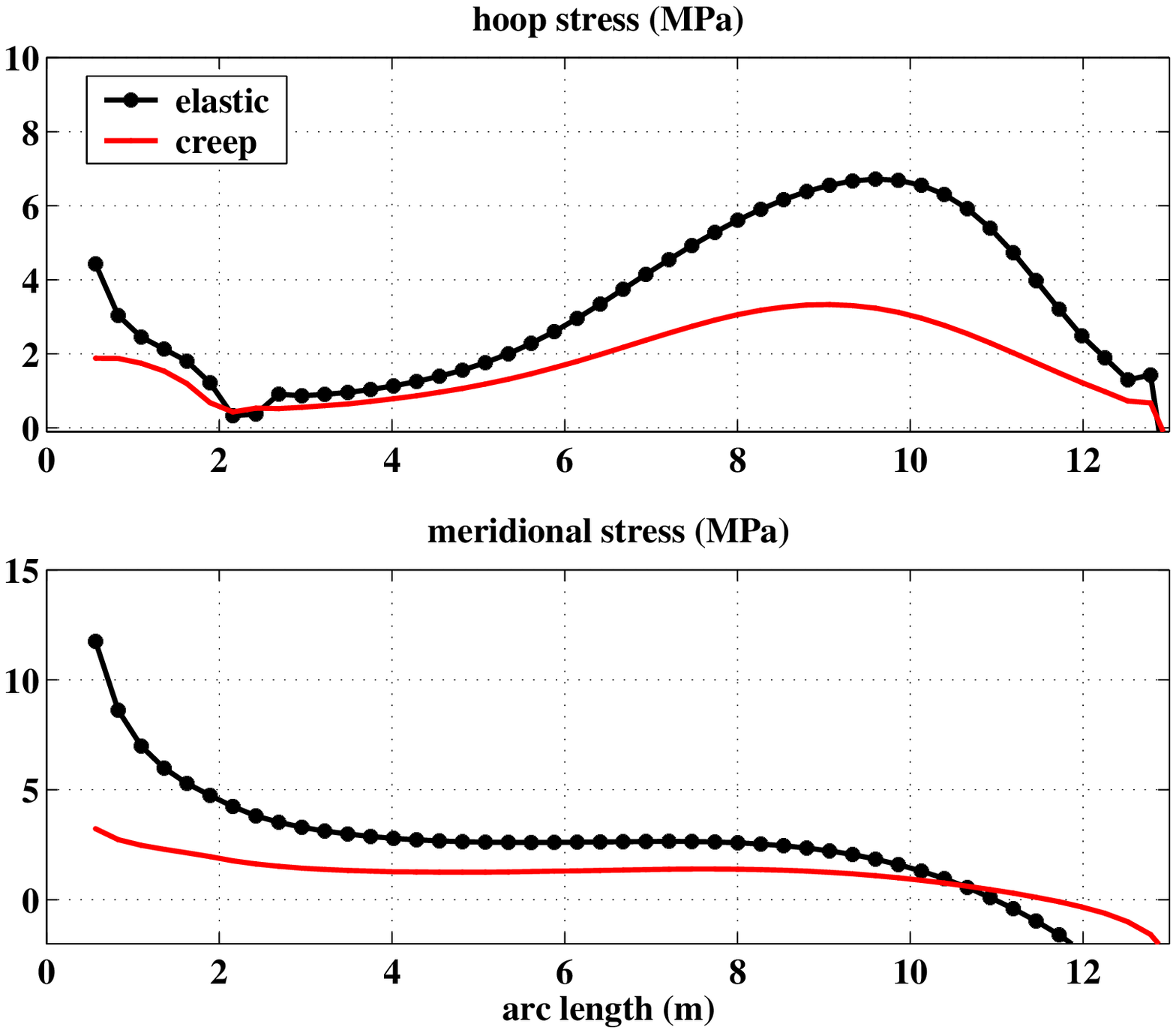,width=2.75in} \\
\multicolumn{2}{c}{ 
100\% relative humidity}
\end{tabular}
\caption{Deformations and stresses of the inner nylon vessel in conditions
of a buoyant load at 0.5\% density difference between scintillator and buffer
fluids.  The difference between models using elastic vs.\ viscoelastic
analyses are shown under the two conditions of 20\% and 100\% relative
humidities in the ambient environment.}
\label{f:stressvisco}
\end{figure}

Figure~\ref{f:stressvisco} and Table~\ref{t:viscoresults}
show how the stress profiles change depending upon whether or not creep is
taken into account,
in the two conditions of 20\% and 100\% relative humidity of the environment.
Regions with negative stress 
are those where wrinkling, not properly modeled by the
 viscoelastic analysis, would occur.

The contact between nylon film and rope is the most challenging
task in the model: the rope follows the deformation of the film, but at the same time constrains 
it and the way this interaction is modeled has important consequences on the analysis results.
Static friction between film and rope attenuates the buoyancy-induced shape deformation and
reduces the stress divergency at the bottom end-plate.
We do not have a measurement of the static friction coefficient between 
film and rope in presence of water or pseudocumene\footnote{The Tensylon datasheet reports $\mu \approx 
0.1$ for the ropes; values for generic nylon in the literature are in the 0.4--0.7 range. 
The datasheet for Capran film (physically similar to Sniamid and Capron) has values between 0.5--0.75 in some cases, or up to 1 in others 
(film-to-film coefficient of friction).}, so neglecting friction is
the conservative thing to do. 
If we assume a static coefficient of friction of 0.5
in our worse-case scenario (0.5\% density difference, wrinkling, no creep),
the hoop stress in the upper hemisphere remains in the 9--10\,MPa range, but the meridional
stress divergence at the south pole drops from 27\,MPa (no friction) to 9\,MPa.

\begin{figure}[h!]
\begin{center}
\epsfig{figure=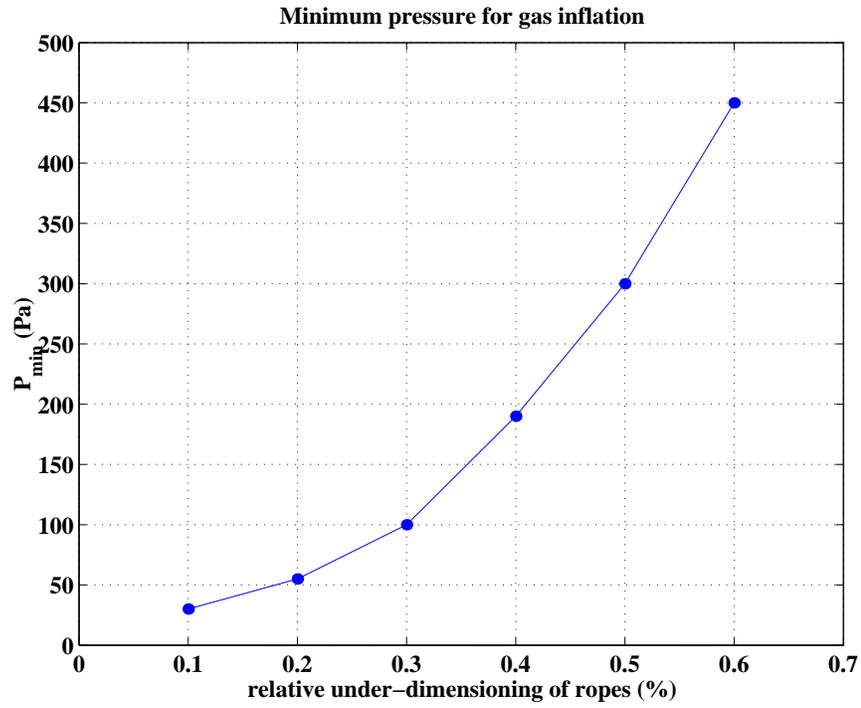,width=4.5in}
\end{center}
\caption{Minimum pressure required, during the gas inflation, 
to compensate the under-dimensioning of the rope and avoid wrinkles.}
\label{f:pminrope}
\end{figure}

\begin{figure}[h!]
\begin{center}
\epsfig{file=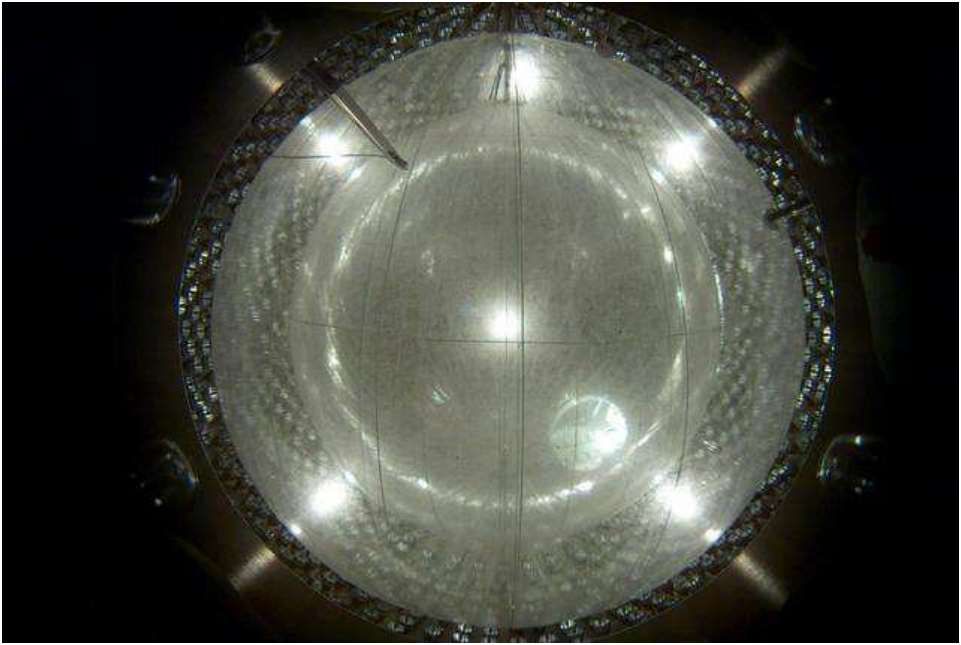,width=5in}
\caption{The nylon vessels in the SSS after inflation with synthetic air.}
\label{f:vessels-inflated}
\end{center}
\end{figure}

The rope length is also a factor in the final stress pattern.
If the ropes are  slack, the stress is larger, as the ropes allow for a 
larger vessel displacement than design, but the difference is only 
visible at the bottom end-plate: in the viscoelastic analysis,
if the rope is 0.35\% too long (9.5\,cm), the equilibrium is reached at about 
10\,MPa instead of 6\,MPa. If the rope is 
0.35\% shorter than design, the level is 5\,MPa. 
From these results, it appears to be convenient to 
pre-stretch, or under-dimension the ropes by some small amount,
compatible with minimizing wrinkles at the bottom end-plate during inflation.
Figure~\ref{f:pminrope} shows the minimum pressure required, at inflation, 
to avoid wrinkles if the rope is initially 
under-dimensioned by a certain percentage amount. 
Friction was not included in these calculations;
this is purely a length conservation issue. The plot hints that 
we can afford a 0.2\% undersize of the rope if, during filling, the
vessel is maintained at a 50\,Pa overpressure.  (In fact, the vessel
ropes {\it were} made slightly undersized, as can be seen by the visible
scalloping of the film in Figure~\ref{f:vessels-inflated}, and the overpressure
is being maintained at this 50\,Pa [5\,mm H$_2$O] value.)

The conclusion we can draw from the FEA model is that if we account for the
creep of nylon under stress and if the hold-down ropes are not over-sized, 
even in the conservative assumption of no film-rope static friction,
in the worst-case scenario the membrane stress on the vessel is 
at or below 10\% of the nylon tensile yield.
In practice, we expect about 5 times lower buoyancy, due only to the
difference in density between scintillator and buffer fluids at the same
temperature.  (The OV, which separates two volumes of buffer
fluid, would experience no buoyant stresses at all.)
Thus we can consider the nylon vessels structurally safe for operation.

\begin{table}[htb]
\caption[]{Fixed-volume vessel filled to 50\,Pa  prior to buoyancy onset,
 no wrinkling, with and without creep 
(Using creep compliance measured for 5\,MPa at 20\% relative humidity,
and C38F data at 7\,MPa from 1992~\cite{cadonati})}
\begin{center}
\begin{tabular}{l|ccc|cc|ccc}
\hline\hline
  & \multicolumn{3}{c|} {$\Delta$R} & \multicolumn{2}{c|} {$\Delta$P} & \multicolumn{3}{c} {Stress}  \\
  & min & max & mean & bottom & top & Meridional  & Hoop &  Rope       \\
 & (cm) & (cm) & (cm) &   (Pa) & (Pa) & (MPa) & (MPa) & (MPa) \\
\hline
\multicolumn{9}{l} {20\% R.H. -  Density gradient: 0.1\%} \\
\hline
Elastic & -5  & 3.5 & 2.2 & 82 & 156 & 5.9 & 2.3 & 13 \\
Creep   & -15 & 11  & 6.4 & 57 & 131 & 2.1 & 1.5 & 14 \\
\hline
\multicolumn{9}{l} {20\% R.H. -  Density gradient: 0.5\%} \\
\hline
Elastic & 16.5& 13 & 8   & 239& 607 & 22.3& 8.5 & 57  \\
Creep   & 31  & 25 & 15  & 153& 521 & 5.4 & 4.4 & 67  \\
\hline
\multicolumn{9}{l} {100\% R.H. -  Density gradient: 0.1\%} \\
\hline
Elastic & 8  & 5  & 3.3  &  76& 150 & 3.4 & 2.0 & 14  \\
Creep   & 18 & 14 & 8    &  51& 125 & 1.4 & 1.3 & 15  \\
\hline
\multicolumn{9}{l} {100\% R.H. -  Density gradient: 0.5\%} \\
\hline
Elastic & 22 & 18 & 11  & 203 & 571 & 12  & 6.7 & 63 \\
Creep   & 36 & 29 & 17  & 129 & 497 & 3.2 & 3.3 & 68 \\
\hline
\end{tabular}
\end{center}
\label{t:viscoresults}
\end{table}

\section{Shipping and installation in Gran Sasso}
\label{s:shipping-installation}

\subsection{Packaging and shipping}
\label{ss:packaging}
To prepare for shipping the vessels, water was inserted at each of the two ends
into the outer vessel (OV), to keep the vessels moist and the air humid.  Then
the nested vessels were triple-bagged: two layers of polyethylene film and a
third aluminized film layer were wrapped around the vessels and heat-sealed.
The aluminized film prevented diffusion of radon into the vessels.

\begin{figure}[h!]
\begin{center}
\epsfig{file=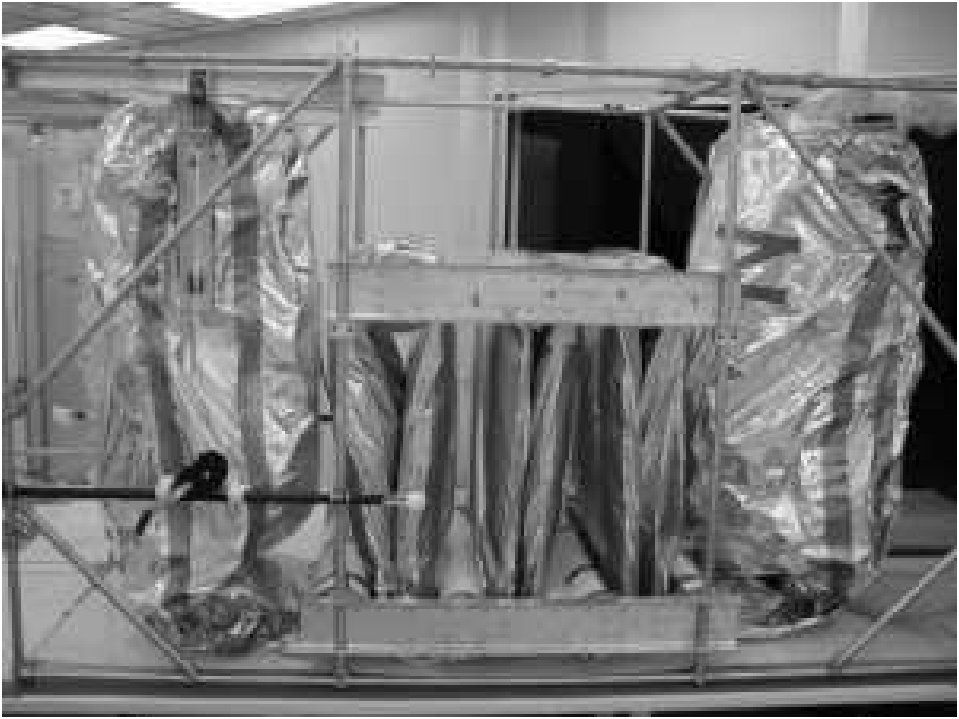,width=5in}
\caption{The set of nylon vessels packaged in the frame for shipping to LNGS.}
\label{f:vessels-package}
\end{center}
\end{figure}

After the sealing of the envelope, the vessels were folded into a packing frame
of approximate dimensions 2.2\,m\,$\times$\,4\,m.  The frame held the two end
regions vertically.  The nylon envelope was folded in the central region of the
frame, as shown in Figure~\ref{f:vessels-package}.  The frame was designed to
fit in a wooden crate at a standard airline cargo size.  The crate was
instrumented with accelerometers and tilt meters to monitor the stress to which
the vessels were subject during the shipping.  The crate was flown from Newark,
New Jersey, USA to Milano, Italy, then trucked into LNGS.

The feed-throughs (section~\ref{sss:break-out}) were added to the vessel end
regions after they had been shipped to LNGS.

\subsection{Installation of the nylon vessels}
\label{ss:installation}
To install the nylon vessels, the north and south end regions needed to be
attached at the top and bottom of the stainless steel sphere (SSS).
Installation was a challenge
because of the weight of the vessel assembly, the 13\,m length of the film
panels, and the large and delicate end regions.  Furthermore, this work had to
avoid damaging the phototubes inside the SSS, which were already installed
everywhere except for a small region on the floor.  The vessels had to be kept
supple throughout the installation, so they were continuously sprayed with
deionized water and the humidity in the SSS was kept $ > 60$\%.

After their arrival at LNGS, the nylon vessels were unpacked in a clean room
connected to the SSS entrance.  Scaffolding was built in the clean room and
SSS to keep the vessels off the floor and then lift them in a controlled
manner.  The vessels were first suspended horizontally from an I-beam track in
the clean room.  They were moved along the track up through the 3\,m port
into the SSS.  Motion was controlled with a system of winches and pulleys.
Once the north end region entered the SSS, it was rotated vertically and
raised to the top, 14\,m above.  The nylon film was unfolded
as it entered the SSS, following the north end region.  While the north end
region was raised, the three layers of bagging protecting the film were
cut away. 

Once the north end region was attached, the film could not simply be allowed to
hang because it was too long.  The length of the film panels when lying flat 
is a factor of $\pi/2$ longer than the
final diameter of the spherical inflated vessels.  Excess film was stored on
a platform on top of the scaffolding in the SSS.  When enough film had been
gathered, the south end region was hoisted upward, rotated vertically, then
lowered and attached to the bottom of the SSS.

\begin{figure}[h!]
\begin{center}
\epsfig{file=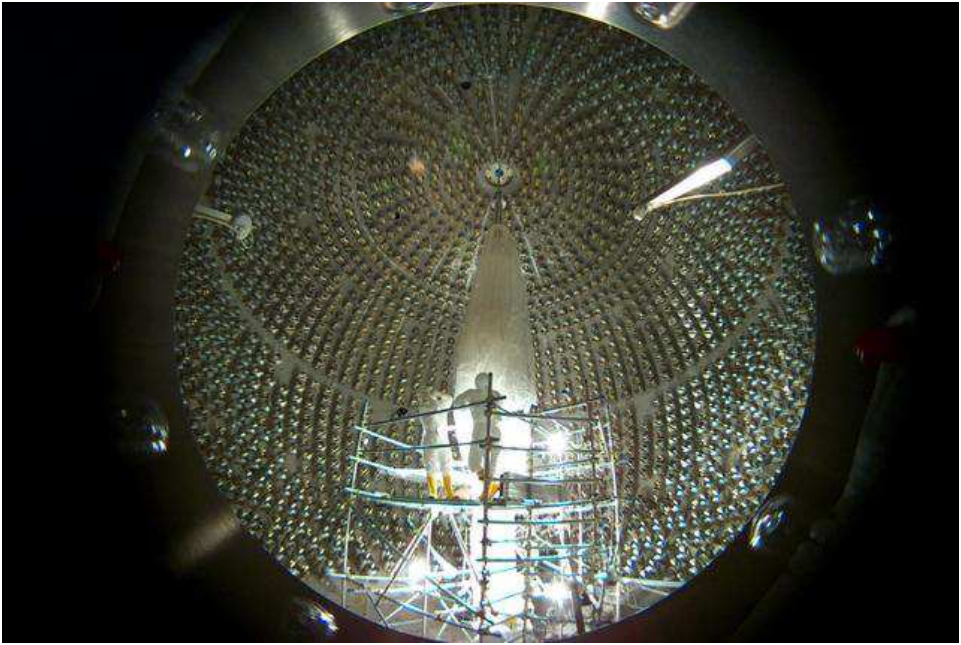,width=5in}
\caption{The nylon vessels partway through the installation.}
\label{f:vessels-installation}
\end{center}
\end{figure}

A photograph of the nylon vessels partway through the installation is
shown in Figure~\ref{f:vessels-installation}.

\subsection{Inflation with synthetic air}
\label{ss:inflation}
Like the initial installation, inflation of the vessels is complicated by the
fact that, in their folded state, the vessels were a bit under $\pi$/2 times
longer than the diameter of the SSS.  The excess film was held, still
folded, in the carriage used for installation.  During inflation, gas was
introduced from the top, so that an expanding inflated region of the vessels
pulled film from the carriage.  Unfolding of the film was controlled by hand
from the installation scaffolding, which was disassembled as the inflation
proceeded.  At all times, the vessels were kept in a humidified, non-brittle
state (as described above) by maintaining $>$ 80\% relative humidity.
Figure~\ref{f:vessels-inflated} shows the fully-inflated nylon vessels.

\subsubsection{Radon limits on synthetic air}
\label{ss:radon-tests}

The vessels were inflated using humidified ultra-low radon air (synthetic
air was used rather than pure nitrogen, in order to protect workers
inside the nearly-sealed SSS from suffocation in the event
of a vessel rupture.)  The synthetic air was produced by mixing high purity
boil-off nitrogen from the Borexino nitrogen plant~\cite{n2-plant}
with oxygen that had been aged in
clean stainless steel bottles to let its radon decay.

The lines used to carry the synthetic air, part of the Borexino permanent
plants, were all made out of electroplated stainless steel. They were
extensively purged with clean nitrogen, and the particulate level at the
delivery point was measured to be better than class 1. The radon activity of
synthetic air injected inside the vessels was continuously monitored, after it
had gone through the inline humidifier, by diverting a small amount from the
main flow and bleeding it through a large electrostatic radon
detector~\cite{kiko}.  Radon levels were constantly below the
$\sim$1\,mBq/m$^3$ sensitivity limit of the device.

For Rn, the main danger is that its daughters wash or leach from the vessels
into the scintillator.   Assuming a worst-case scenario of all daughters
transported into the liquid, a Rn concentration of $\leqslant$ 0.1\,mBq/m$^3$
during inflation will lead to $<$ 1~count/day of all 210-chain daughters in
the fiducial volume.

\subsubsection{Leak rate tests}
\label{ss:leak-rate-tests}

Both vessels were leak checked after inflation. SF$_6$ tracer gas, aged to let
radon decay away, was injected into the inner vessel (IV) air stream at
$\sim$1\% concentration. The leak checking strategy relied on measuring the
SF$_6$ concentration in the outer buffer volume (between IV and OV) as a
function of time; with the IV over-pressure and the volume of the outer buffer
it would yield the total IV leak rate. This technique turned out to be affected
by sampling problems and by the difficulty of mixing the gas in the outer
buffer.  It gave an upper limit of $<$ 5\,cc (gas)/s/mbar, equivalent to a
liquid leak rate of 0.05\,cc (pseudocumene)/s/mbar.

A more sensitive method, applicable to both vessels, was to simply monitor the
over-pressure trends of the IV and OV.  After correcting for external
atmospheric pressure fluctuations (the SSS was still open to the outside at
this stage) and with a couple of months of data the following gas leak rates
were recorded:
\begin{itemize}
\item $\sim$ 0.5\,cc (gas)/s/mbar, or 0.005 cc (pseudocumene)/s/mbar, for the IV
\item $<$ 10\,cc (gas)/s/mbar, or 0.1 cc (pseudocumene)/s/mbar, for the OV.
\end{itemize}
The value for the IV is a factor of two better than the design specifications
specified in section~\ref{ss:leak-tightness}, and has to be taken as a conservative number in that the elastic deformations and creep of the membrane were not taken into account. The leak rate of the OV is at least one order of magnitude better than required.

\subsection{Pressure control and DCS in filling stations}
\label{ss:dcs}
Because the vessels are among the most fragile parts of the Borexino
experiment, it is necessary to monitor their volumes, pressures, and shapes
continuously.  If one or more of these parameters departs from the desired
limits, it is necessary to take action to bring it back into the acceptable
range.  Borexino has a unique digital control system (DCS) to monitor and
control all the process plants.  In addition to showing the readouts of the
strain gauges and temperature sensors on the vessels (refer to
sections~\ref{sss:load-cells} and~\ref{sss:temperature-sensors}, respectively),
the DCS was also used in the initial gas inflation of the vessels, and to
ensure that the vessel parameters remained in a steady state while waiting to
begin the process of filling the detector with water.

Since the atmospheric
pressure in the laboratory may vary freely, the crucial parameters for these
operations are the {\it differential} pressures across the IV and
across the OV.  These values are measured using differential pressure
(dP) cells connected to the DCS.  The dP cells are the Fisher Rosemount
model 3051C.  They are highly accurate, to 0.04\% of their full range.

During the initial inflation, automatic interlocks between the dP cells and the
valves used for gas insertion were used: in case a value went higher then the
set threshold, if the interlock was enabled, the connected valves would close,
stopping the inflation.  The total value of the strain gauges gave a good
indication of the inflation state.  (The strain gauges and the temperature
sensors will also be critical in monitoring the vessels during the filling with
water and pseudocumene.) Other useful parameters monitored during the inflation
included the relative humidity (this was kept high in order to keep the nylon
film supple) and the oxygen content of the synthetic air.

For many months, the vessels remained inflated with synthetic air.
During these static conditions, as well as during the following purges with
nitrogen, critical high-level alarms are set on the dP cells via the DCS.  On
the basis of their readouts, actions to re-inflate or deflate one or both
vessels may be taken.

In addition to the DCS, the vessels have a passive system for control of
differential pressures, consisting of two bubblers.  By changing the levels of
the water in these, it is possible to set the maximum differential pressure
allowed between the three volumes. The idea is to use this passive system as a
last resort to save the vessels in case of sudden drastic changes.

\subsection{Purging with nitrogen}
\label{ss:purging}

Normal air, even when dust-free, has an activity of several Bq/m$^3$ due to the
presence of radon and other radioactive noble gas isotopes.  Even one cubic
centimeter of air thus implies about one radioactive decay in the fiducial
volume per day.  It is therefore crucial to purge the inflated vessels with
ultra-pure nitrogen before filling them with water or scintillator.

The purging gas used in the vessels has very stringent radiopurity
requirements.  The primary concerns are $^{222}$Rn and the beta emitters
$^{39}$Ar (565\,keV endpoint, 369\,yr half life), present in natural Ar at a
level giving $\sim$ 1.2\,Bq/(stp m$^3$ Ar), and  $^{85}$Kr (687\,keV endpoint,
10.8\,yr half-life), present in natural Kr at a level giving $\sim$ 1\,MBq/(stp
m$^3$ Kr).   Kr and Ar will diffusively or convectively be transported into the
water during water filling, and then to the pseudocumene during water displacement by pseudocumene.
Rn introduced with the gas will quickly decay, leading to $^{210}$Pb, Bi and Po
deposited on the surfaces of the vessels.

The amount of Kr and Ar transfered to the scintillator from the purge gas
depends on the relative solubilities of these gasses in N$_2$, water and pseudocumene,
and the amount of mixing or diffusion.    Conservatively we assume full
equilibration between Kr and Ar in the gas/water phases during water filling
and subsequent full equilibration of Kr and Ar between pseudocumene and water during pseudocumene
displacement of the water.  The relevant solubilities are shown in
Table~\ref{t:solubilities}.
With a goal of $\leqslant$ 1\,count/day in the fiducial volume, we arrive at
required (volume, or molar) concentration limits of 0.37\,ppm for Ar,
and $\sim 0.1$\,ppt Kr in the purge gas.

\begin{table}
\caption{\label{t:solubilities}Relative molar solubilities
(volume of gas dissolved per unit volume
of liquid) of Ar, Kr and Rn in water and pseudocumene at 15$^\circ$C.  By
definition, the solubility of each in N$_2$ gas is 1.  For
example, in equilibrium the molar concentration of Ar in N$_2$ gas is $1/0.039
= 26$ times its concentration in water.  Data taken from 20$^\circ$C values
in the Solubility Data Series, extrapolated to 15$^\circ$~\cite{ar-sol,rn-sol}.
The estimated values for pseudocumene are averages of available data for the
similar compounds benzene and toluene (which differ by $\sim$ 20 \%).}
\begin{center}
\begin{tabular}{lrr}
\hline\hline
Gas     & water & pseudocumene (est.) \\\hline
Ar      & 0.039 & 0.23 \\
Kr      & 0.073 & 0.77 \\
Rn      & 0.295 & 12.96 \\ \hline
\end{tabular}
\end{center}
\end{table}

To start with, the SSS was filled with normal air, while the nylon vessels
(IV interior and inner buffer region) had been inflated with
synthetic air.
During this time period, Kr and Ar had diffused into
the nylon film and tubes of the vessels; the volume of these gases trapped in
these plastic parts was equivalent to a few liters of air.  Purging therefore
was done in stages, with a few weeks between each stage, to allow the Kr and
Ar to diffuse back out.  The first several stages, amounting to five volume
exchanges within the IV, were done with high purity
nitrogen (HPN) with a concentration of a few ppm Ar.  The purging was
finished with 14 volume exchanges in the IV of special low Ar and
Kr nitrogen (LAKN).

During purging, the relative humidity of the nitrogen was maintained at
over 60\% to preserve the flexibility of the nylon vessels.  Flow rate of
the incoming gas was about 25\,m$^3$/hr for the inner buffer, and
30\,m$^3$/hr for the other two volumes.  Differential pressure between the
vessels was kept below 5\,mm\,H$_2$O at all times.

\begin{figure}[h!]
\begin{center}
\epsfig{file=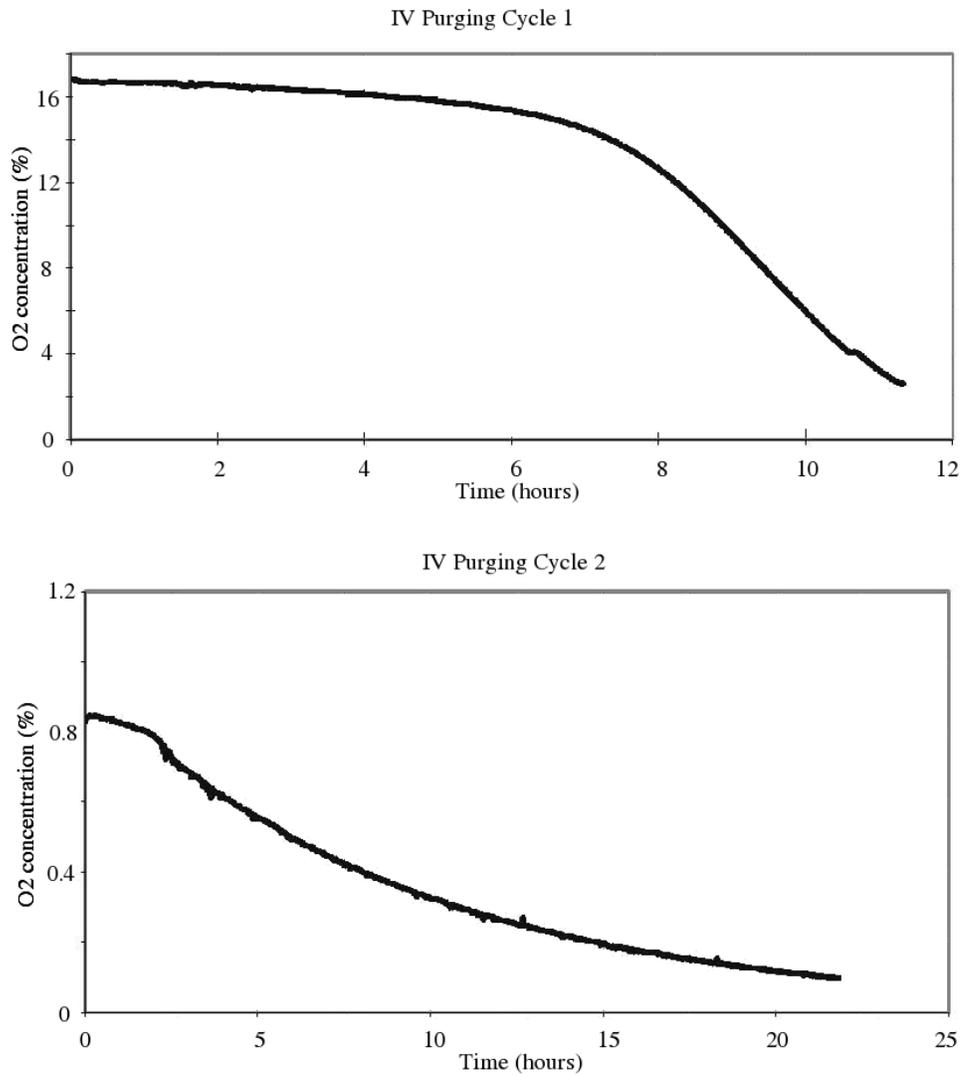,width=5in}
\caption{Graphs of the O$_2$ concentration in the IV during the first
two stages of purging.  Time in hours is shown on the horizontal, and the
percentage of O$_2$ on the vertical.  After the end of the first purging
stage, the stratified gas continued to mix, so the concentration is lower
at the beginning of the second stage.}
\label{f:purging}
\end{center}
\end{figure}

Concentrations of the relevant gases, O$_2$, Ar, SF$_6$ (used as a tracer), and
N$_2$, in gas leaving the vessels at the bottom were monitored at 5\,s
intervals with a residual gas analyzer.  At the end of each purging stage, more
accurate measurements, scanning all mass numbers, were taken.  During the first
stage (top of Figure~\ref{f:purging}), the reduction factor in the IV was about
20, much higher than the expected per-volume factor of e (2.718).  This
is a result of stratification in the vessel, with the pure lighter N$_2$ coming
in from the top and denser air exiting at the bottom.  In following stages, the
reduction factor per volume was much closer to e with
complete mixing, shown in the bottom of Figure~\ref{f:purging} (the second
purging stage was about 2 volumes).

\section{Summary of Radiopurity}
\label{s:radiopurity}

In a low background experiment such as Borexino, each detector component is
subject to strict radiopurity requirements.  The nylon film for the inner
vessel (IV), in particular, is a potential background source from
$\gamma$~emission (external background) and from Rn emanation into the
sensitive volume (internal background).  Other structural components of the
detector also release $\gamma$ rays and Rn, although most of the externally
produced Rn will be barred from entering the volume of scintillator by the two
nylon vessels.

\subsection{Gamma backgrounds}

The design requirement for the radiopurity of materials used in the vessel
fabrication is that the nylon vessels and their associated parts should produce
a $\gamma$ background rate less than the $\gamma$ background from the
photomultiplier tubes and light cones mounted on the stainless steel sphere
(SSS).  (Though the selected phototubes incorporate special low-radioactivity
Schott 8246 glass, the glass nevertheless has concentrations of 30\,ppb U,
10\,ppb Th, and 20\,ppm K by mass, resulting in the phototubes
being among the main sources of external $\gamma$ rays.)  Moreover, the overall
$\gamma$ background in the neutrino window should be less than the neutrino
rate.

The radius of the IV is 4.25\,m, while the radius of the fiducial
volume is 3\,m. Thus, there is an active shield of 1.25\,m between the vessel
and the part of the detector from which neutrino event data will be used.  With
this shield, the maximum allowed concentrations of U, Th  and K in the nylon
film are 10\,ppt, 20\,ppt and 30\,ppb, respectively.  Such limits guarantee
that the external background from the IV film is less than 10\% of
the external background from the photomultiplier tubes.  The thin, flexible
IV membrane has a total mass of only 32\,kg.  We note that the use of
a rigid shell instead, which by necessity would be much more massive, as a
scintillator containment vessel in Borexino would have required much more
stringent limits on the permissible concentrations of $\gamma$ emitters.

The materials used in the construction of the outer vessel (OV) end regions
(see Table~\ref{t:EndRegionRad}) are the principal sources of vessel-related
$\gamma$ background.  Table~\ref{t:GammaBackg} reports the simulated $\gamma$
background from each end region component in the energy windows for $^7$Be
neutrinos (250--800\,keV) and $pep$ neutrinos (800--1300\,keV) in different
fiducial volumes.  The steel flanges and the steel pipe located half-way
between IV and OV contribute the most, due to the $^{60}$Co content in steel.
As noted above, these numbers are conservative because the $^{60}$Co half-life
is only 5.3~years.

\begin{figure}[h!]
\begin{center}
\epsfig{file=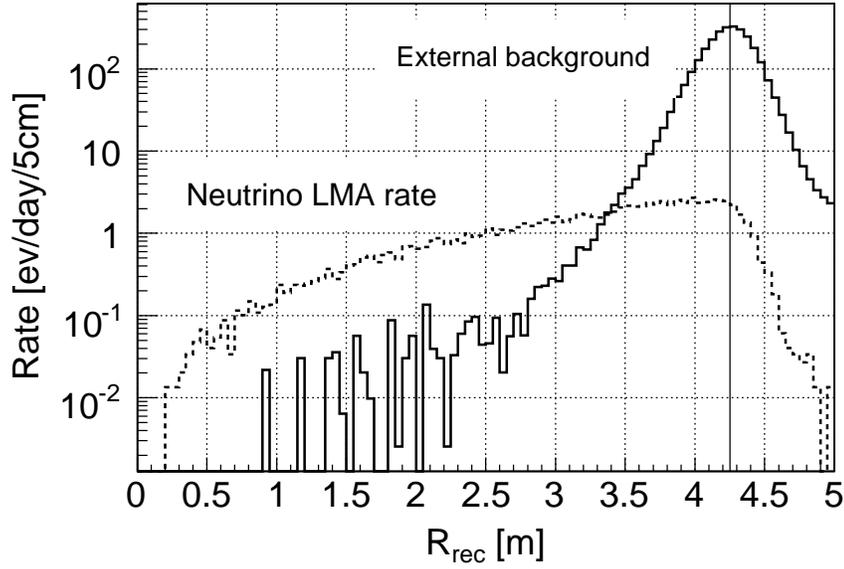, width=5in}
\caption 
{ Radial distribution of the {\it reconstructed} positions of external
$\gamma$-ray and neutrino scintillation events in the energy window
0.25--0.8\,MeV.
Note that signal-to-noise remains greater than one in all radial shells
of scintillator only for a fiducial volume of radius $<$ 3.3\,m or so,
corresponding to a scintillator mass of $<$ 130~tons.
The vertical bar at 4.25\,m is the position of the inner nylon vessel.
Since this simulation takes the finite resolution of position reconstruction
into account, some events have a reconstructed position
outside the inner vessel.}
\label{f:radial}
\end{center}
\end{figure}

For ease of comparison, Table~\ref{t:GammaBackg} shows the predicted neutrino
rate and the expected background from the end region components,
nylon membranes, phototubes, their light cones, and the SSS.  The table was
constructed using the
GENEB (GEneration of NEutrino and Background interactions) Monte Carlo software
developed expressly for simulation of the Borexino detector~\cite{geneb},
including a simulation of the finite resolution of position reconstruction.
The radial dependence of neutrino and external
background events in the 0.25--0.8\,MeV energy window can be seen in
Figure~\ref{f:radial}: the different behaviors make the fiducial volume cut a
powerful tool to suppress this type of background.

\begin{table}[p!]
\caption{\label{t:GammaBackg}Gamma background in Borexino
from the various components of the end region. The 
background is expressed as rate in events/day in the 250--800\,keV
energy window (NW) and in the 800--1300\,keV energy window (pep).  Rates are
given for the entire volume of scintillator (``all''), and for 200-ton and 100-ton
central fiducial volumes.}
\begin{center}
\small 
\begin{tabular}{llr|rr|rr|rr}
\hline\hline
Component & Position & Mass   & \multicolumn{2}{c|}{All} &
                                \multicolumn{2}{c|}{200\,t} &
                                \multicolumn{2}{c}{100\,t} \\
          &   [m]  & [kg] & NW & pep & NW & pep & NW & pep \\
\hline
nylon rings          & 4.25 & 5.2  & 124  & 107  & 2.16  & 1.53  & 0.15  & 0.180\\
nylon hubs           & 4.25 & 0.4  &  12  &   8  & 0.43  & 0.14  & 0.03  & 0.008 \\
Cu rim brackets      & 4.35 & 1.6  &  32  &  17  & 0.53  & 0.25  & 0.03  & 0.025 \\
nylon tubes     & 4.25--4.9 & 1.9  &  31  &  27  & 0.73  & 0.45  & 0.14  & 0.060 \\
Cu struts       & 4.25--4.9 & 11.2 &  56  &  21  & 1.21  & 0.40  & 0.23  & 0.034 \\
nylon flanges       & 4.9   & 0.6  &   1  &   1  & 0.04  & 0.04  & 0.01  & 0.009 \\
steel flanges       & 4.9   & 3.4  & 63   & 29   & 1.47  & 1.53  & 0.30  & 0.061 \\
steel tubes       & 4.9--5.6& 6.9  & 66   & 30   & 2.52  & 1.70  & 0.68  & 0.176 \\
IV rope attachment   & 5.45 & 3.7  &  4   &  1   & 0.14  &  0.09 & 0.02  & 0.002 \\
steel OV mount rings & 5.5  & 7.8  & 13   & 6    & 0.32  &  0.31 & 0.06  & 0.011  \\
steel OV tubes       & 5.5  &  2.6 &  2   & 1    & 0.06  &  0.06 & 0.01  & 0.003  \\
steel OV flanges     & 5.6  & 11.3 &  6   & 2    & 0.23  &  0.16 & 0.03  & 0.004  \\
4 pipe load cells    & 5.2  &  0.3 &  3   & 2    & 0.09  &  0.09 & 0.02  & 0.003  \\
12 string load cells & 5.4  &  0.8 &  2   & 1    & 0.06  &  0.06 & 0.01  & 0.002  \\
temperature sensors &4.9--5.6& 0.2 &  9   & 4    & 0.39  &  0.09 & 0.04  & 0.003  \\
\multicolumn{3}{l|}{\it Subtotal, End Regions}
                                   & 424  & 257  & 10.38 & 6.90  & 1.76  & 0.581  \\
\hline
IV nylon film        & 4.25 & 32.1 &   52 &  60  &  0.39 & 0.34  & 0.01  & 0.006  \\
optical fibers       & 4.25 & 0.5  &   15 &   9  &  0.12 & 0.07 &  0.01  & 0.001  \\
ropes                & 4.25 & 4.5  &  332 & 180  &  2.45 & 1.58  & 0.06  & 0.028  \\
\multicolumn{3}{l|}{\it Subtotal, Nylon Vessels} 
                                   &  823 & 506  & 13.34 & 8.89  & 1.84  & 0.616 \\
\hline
light cones       &6.3--6.5 &      &  896 & 494  & 18    & 18    &  0.6  & 0.4 \\ 
phototubes        &6.5--6.85&      & 1280 & 651  & 25    & 26    &  0.6  & 0.4 \\
SSS               &6.85&           &  496 & 207  & 10    &  9    &  0.2  & 0.2 \\ 
\it{Total, Borexino}          & &  & 3495 & 1858 & 66    & 62    &  3.2  & 1.6 \\
\hline
Neutrino rate        &$<$ 4.25 &   &  89  & 4    & 64    & 2.8   &  32   & 1.4 \\
\hline
\end{tabular}
\end{center}
\end{table}

\subsection{Radon diffusion through nylon}
\label{ss:rn-diffusion}

A more stringent radiopurity requirement is the amount of radon allowed to
diffuse into the scintillator before it decays into $^{218}$Po and succeeding,
essentially non-diffusing, daughters.
The diffusion of radon through nylon films, if known, may be used to evaluate
the potential background from radon originating in the bulk nylon due to
$^{226}$Ra impurity in the material, and due to radon that originates in the
buffer region and diffuses through the membrane into the scintillator. 

The passage of radon through nylon, regardless of whether a Rn atom originates
within the IV nylon film and travels into the scintillator, or
instead travels from the outside to the inside of the IV, is governed
by the diffusion equation, with two additional terms accounting for the
radon decay and for any radon produced within the nylon:
\begin{equation}
\frac{\partial\rho}{\partial t} = D \nabla^2 \rho - \frac{\rho}{\tau} +
\mathcal{A}.
\end{equation}
Here, $\rho$ is the local radon concentration at a point within the film,
$D$ is the diffusion coefficient of radon in nylon, $\tau$ the
radon mean lifetime of 5.5~days, and $\mathcal{A}$ the production rate
of radon (or equivalently, decay rate of radium) per unit volume of nylon.

The problem is essentially one-dimensional when, as in Borexino, the vessel
radius is much greater than the membrane thickness $d$.
We may presume a steady-state
solution.  An important length
scale, giving the mean distance that a radon atom travels
(projected onto the radial direction) through the nylon film before it decays,
is
\begin{equation}\label{e:difflength}
  \ell=\sqrt{D\tau}.
\end{equation}

\subsubsection{Permeability of nylon to radon}
\label{sss:nylon-permeability}

The permeability $P$ of non-decaying atoms through a membrane is defined 
as the product of the solubility $S$ (the ratio of concentrations of a solute
at the interface between two solvent materials) with the diffusion coefficient, 
which regulates the speed of propagation through the membrane material. 
In order to take into account the decay of $^{222}$Rn as it 
crosses the nylon membrane, we define an effective permeability, 
which depends on the membrane thickness $d$ and on the
diffusion length $\ell$, which was defined in eq.~\ref{e:difflength}:
\begin{equation}\label{e:Peff}
P_{\rm eff}=D S \frac{d/\ell}{\sinh(d/\ell)}.
\end{equation}
In this particular case, the solubility factor $S$ is defined as the ratio of
the radon concentration
in the fluid touching the nylon to the radon concentration in the nylon
itself at the surface.

\begin{table}[t]
\caption{\label{t:RnDiffHumidity}Measured diffusion properties 
of $^{222}\mbox{Rn}$ in nylon as functions of relative humidity (RH).
Data from~\cite{cadonati,rn-humidity}; see text for details.
The top two lines are for nylon film between two volumes
of dry pseudocumene (PC) and for nylon separating volumes of pseudocumene
and water.  The diffusion length $\ell$
is defined in eq.~\ref{e:difflength}; the effective permeability $P_{\rm eff}$
is defined in eq.~\ref{e:Peff} and computed for the thickness of the Borexino
IV, $d=125\,\mu$m.}
\begin{center}
\small 
\begin{tabular}{lrrccccc}
\hline\hline
interface & RH   & d  & $D$ & $S$  & $\ell$ & $P_{\rm eff}$ \\
& $\left[ \% \right]$ & [$\mu$m]& [cm$^2$/s] &   &   [$\mu$m] &  [cm$^2$/s] \\
\hline
PC/PC  &$\sim 0$ & 15 & -- & -- & -- & $<2\times 10^{-12}$\\
H$_2$O/PC &100 & 135 & $(2.98\pm 0.15)\, 10^{-10}$ & $13\pm 2$& 120 & $(3.2\pm 0.6)\, 10^{-9}$\\
\hline
N$_2$/N$_2$&$\sim 0$ &18&$(2.1\pm 0.4)\, 10^{-12}$ & $4.5\pm 0.7$ & 10  & $(8.8\pm 2.3)\, 10^{-16}$  \\
N$_2$/N$_2$&$12$     &18&$(2.2\pm 0.3)\, 10^{-12}$ & $2.5\pm 0.3$ & 10  & $(5.9\pm 1.2)\, 10^{-16}$  \\
N$_2$/N$_2$&$32$     &18&$(4.3\pm 0.5)\, 10^{-12}$ & $1.8\pm 0.2$ & 14  & $(2.2\pm 0.4)\, 10^{-14}$  \\
N$_2$/N$_2$&$52$     &18&$(1.9\pm 0.3)\, 10^{-11}$ & $1.4\pm 0.2$ & 30  & $(3.5\pm 0.8)\, 10^{-12}$  \\
N$_2$/N$_2$&$88$     &18&$(1.3\pm 0.2)\, 10^{-10}$ & $1.5\pm 0.2$ & 79  & $(1.3\pm 0.3)\, 10^{-10}$  \\
N$_2$/N$_2$&$100$    &18&$(1.3\pm 0.2)\, 10^{-9}$  & $0.7\pm 0.1$ & 248 & $(8.7\pm 1.9)\, 10^{-10}$  \\
\hline
\end{tabular}
\end{center}
\end{table}

The diffusion constant $D$ in nylon depends strongly on the relative humidity
of the nylon, and increases as the humidity is raised.  It is such that the
diffusion length $\ell$ is $\sim 10\,\mu$m for dry nylon  and $\sim 100\,\mu$m
for wet nylon.  The effective permeability increases likewise.
Table~\ref{t:RnDiffHumidity} reports measured diffusion
coefficient, solubility and the effective permeability of $^{222}$Rn in a
$125\,\mu$m nylon barrier at various relative humidities. 

The first two rows in the table report results obtained in Princeton and
described in reference~\cite{cadonati}. These measurements were conducted with
fluid on both sides of a nylon membrane, in order to reproduce the experimental
conditions for the IV. On one side of the membrane was a small can
with radon-saturated fluid (either pseudocumene or water), on the other side a
larger cylindrical stainless steel tube held radon-free scintillator
(pseudocumene plus fluors), viewed by two phototubes that measured the decay
rate of radon permeating through the membrane, function of time.  In the first
case, with dry pseudocumene on both sides, the diffusion was below the
sensitivity limit for the apparatus, even using the thinnest available nylon
film ($15\,\mu$m nylon-6). The second (wet) measurement was performed with C38F
nylon film of thickness comparable to that of the Borexino IV.

The remaining entries in Table~\ref{t:RnDiffHumidity} are for measurements
performed by our collaborators in Heidelberg and Krak\'ow~\cite{rn-humidity},
in which a $18\,\mu$m nylon C38F membrane was held in a nitrogen atmosphere at
a known relative humidity: on one side the gas contains a known amount of
radon, from a source, and on the other side a photomultiplier tube is used to
detect radon that has permeated through the membrane. In order to control the
humidity, the nitrogen is passed through a two-liter glass flask partially
filled with acqueous solutions of various salts; depending on the salt, it was
possible to obtain several standard humidity values in the gas phase in
equilibrium above the solution.

\subsubsection{Radon emanation from nylon}
\label{sss:rn-emanation}

In general, the concentration of radon within the nylon film will be much
greater than that in the liquid in contact with either inner or outer surface,
as pseudocumene may be purified much more efficiently than nylon.
Therefore, in the limit of a large mean travel distance for a radon atom,
$\ell/d \gg 1$, the
average radon activity within the Borexino scintillator would be one-half
of the total activity within the IV film, scaled by the ratio of
the nylon volume to the scintillator volume.  (The other half would diffuse
outward into the inner buffer volume.)  In particular, this upper limit
corresponds to about 10 radon events per day within a 100-ton fiducial
volume, if we assume complete mixing of the scintillator.  Convection in
Borexino is actually unlikely, due to a fair temperature gradient (a few-degree
increase from bottom to top), so the actual radon activity within
the fiducial volume should be much smaller.

More detailed calculations~\cite{mccarty,rn-humidity} indicate that radon activity
within the 100-ton fiducial volume, due to radon emanation from the IV,
will be between 1--7 radon decays per day, depending upon the
actual relative humidity seen by the nylon film.  Again, these figures
presume complete mixing of the scintillator on a time-scale comparable to
the radon mean lifetime or less.

In the assumption of secular equilibrium, the $^{222}$Rn emanation from
$1\,$ppt $^{238}$U in nylon and emanating into the fiducial volume, in a dry
Borexino detector, is equivalent to the $^{222}$Rn from less than 10$^{-17}$\,g
of $^{238}$U per gram of scintillator.  We are hence fairly confident that
radon emanation from nylon will be a small effect compared to uranium
(and radium) contamination in the scintillator fluid.

\subsubsection{Principle of the outer vessel barrier}
\label{sss:ov-barrier}

The OV was designed to be a radon barrier that would minimize the radon
concentration near the IV. The plan for the OV barrier was based on the concern
that the outer region of the SSS, which includes the phototubes and SSS inner
surface, could contain a significant quantity of dust due to the exposure of
several months required for installation of the photomultiplier tubes.  This is
true even though the space was operated with filtered Class 1000 clean-room
air. Moreover, although this region was cleaned just before the closing of the
SSS, a precision cleaning could not be accomplished due to the limited
accessibility.  A protective barrier was judged to be the best way to assure a
low background from dust and radon at the IV.

The effectiveness of the OV as a radon barrier was evaluated by calculating the
rate of radon diffusion through the nylon membrane~\cite{cadonati}.  The
relevant parameters are the  mean diffusion length $\ell$
(eq.~\ref{e:difflength}), which is the mean distance a radon atom travels
during its mean life $\tau \approx 5$\,days, and the thickness $d$.  The
solution of the diffusion equation for a decaying substance through a barrier,
with the simplifying hypothesis of relatively large concentration of radon
($\rho_{\rm out}$) on one side and negligible concentration ($\rho_{\rm in}$)
on the other, yields the following:
\begin{equation}
\rho_{\rm in} = P_{\rm eff}\frac{\tau}{d^2}\frac{V_{\rm barrier}}{V_{\rm in}}\, \rho_{\rm out}
\end{equation}
where $V_{\rm barrier}=4\pi R^2d$ is the volume of the membrane, in the
Borexino spherical geometry, and  $V_{\rm in}=4\pi R^3/3$ is the enclosed
volume.  If we substitute the definition of $P_{\rm eff}$ from eq.~\ref{e:Peff}
and use the definition of diffusion length from eq.~\ref{e:difflength}, the
equation becomes
\begin{equation}\label{e:barrier}
\frac{\rho_{\rm in}}{\rho_{\rm out}} = \frac{\ell/d}{\sinh(d/\ell)} \frac{3Sd}{R}.
\end{equation}

For wet film, the mean diffusion distance is $\sim 100\,\mu$m, which is
comparable to the thickness of the film ($125\,\mu$m); then $\ell/d$ and $\sinh
(d/\ell) \approx 1$. In this case, the result is consistent with the following
intuitive explanation.  If the membrane thickness is equal to the mean
diffusion distance, then the flux of atoms going into the vessel is the number
of atoms in the membrane divided by the mean life of radon. Since these atoms
decay with the same mean life inside the vessel, the result is that the radon
concentration (atoms/m$^3$) within the vessel is the number of radon atoms in
the barrier divided by the volume of the vessel. This ratio is the
concentration in the membrane times the ratio of the volume of the membrane
($4\pi R^2d$) divided by the volume of the vessel ($4\pi R^3/3$).  The
concentration within the film is equal to the concentration outside the vessel
times the solubility ratio $S$ for radon in pseudocumene
and in nylon.  The final result
is that the concentration inside the vessel is given by
$$\frac{\rho_{\rm in}}{\rho_{\rm out}} = \frac{3Sd}{R}.$$

Consider only the OV for now.
For $d = 125\,\mu$m, $R = 5.5$\,m (the OV radius), it follows that,
for dry nylon, eq.~\ref{e:barrier} yields
$$\frac{\rho_{\rm in}}{\rho_{\rm out}} \approx 2 \times 10^{-10}.$$
In the worst-case scenario of nylon in contact with water, $S \approx 10$, and
$$\frac{\rho_{\rm in}}{\rho_{\rm out}} \approx 10^{-3}.$$
Thus, we find that a thin nylon film will serve as an excellent barrier even
in presence of water.

In Borexino, with two concentric nylon membranes, the reduction factor for
radon concentration between the outer buffer and scintillator volumes is
roughly the square of the reduction factor for a single vessel.  Incorporating
geometric considerations (the intermediate inner buffer region is a shell, not
a full sphere), we find this range for the ratio between radon concentrations
in the scintillator (IV) and outer buffer (OB):
$$10^{-19} \le \frac{\rho_{\rm IV}}{\rho_{\rm OB}} \le 10^{-6}$$
(the number on the left being the dry nylon value and that on the right, for
complete immersion in water).  
Hence, by far the dominant contribution to radon within the scintillator will
be emanation from the IV nylon itself.

As above, all of these formulas presume complete convective mixing of radon
within the scintillator fluid.  A lack of convection will imply that most radon
atoms that manage to emanate from or pass through the IV film nevertheless do
not ever reach the central fiducial volume of scintillator.

\section{Conclusions}
\label{s:conclusions}
The Borexino nylon vessels have been installed within the detector for over
three years.  During this time, members of the Borexino collaboration have
learned a great deal about how to ensure their safety under various operating
conditions.  Water filling of the detector was initiated in August 2006, and
finished 3.5~months later.  Already the detector has observed \v{C}erenkov
light from muons passing sideways through the detector, generated by the
interaction of $\mu$ neutrinos (created at the CNGS facility of
CERN~\cite{cngs}) with the water molecules.

Filling the detector
with purified buffer and scintillator is ongoing in early 2007, to be followed
shortly by filling the outer steel tank (muon detector and neutron shield)
with water.  When this final stage is
finished, the detector will be complete, able to observe the $^7$Be and
possibly $pep$ and even CNO-cycle neutrinos.  Only at this point will we
know the levels of radioactive contamination present within the detector
fiducial volume.  We may say, optimistically, that there is no reason to
believe any unexpected major radiocontamination issues will present themselves,
and every reason to expect that the Borexino nylon vessels will fulfill all the
needed requirements for operating successfully.

\section*{Acknowledgments}
The authors would like to thank the following individuals and companies,
all of whom have been critical in the design, construction, or installation
of the Borexino vessels and development of associated technologies.

John-Paul Chou, Farng-Yi Foo, Charles Sule, Joseph Stritar, and Thomas Zhang
were invaluable in the measurements of mechanical properties of nylon film at
Princeton.  The Borexino vessel construction crew included Julie Bert,
John-Paul Chou, Joel Greenberg, Eric Hopkins, Brian Kennedy, Ted Lewis, John
Saunders, Domenic Schimizzi, and Charles Sule.  Davide Gaiotto was an important
contributor to the research and development effort on the optical fibers.  The
vessel installation crew included Augusto Brigatti, Angelo Corsi, Antonio di
Ludovico, Giuseppe di Pietro, Massimo Orsini, Sergio Parmeggiano, and others.

Evan Variano helped early in the research effort on radon diffusion.  Other
individuals involved in designing, building and operating the radon scrubber
for the vessel construction clean room included Nick Darnton and Costin Bontas.
Aldo Ianni, Matthias Laubenstein, and the late Burkhard Freudiger helped to
monitor radon levels in the synthetic air during the vessel inflation.  Radon
and radium measurements performed by our collaborators at Heidelberg and
Krak\'ow, including Christian Buck, Gerd Heusser, Wolfgang Rau, Hardy Simgen,
Wojciech Wlaz{\l}o, Marcin W\'ojcik, and Grzegorz Zuzel, were invaluable.

Mathematical calculations of stress on the Borexino vessels came in great part
from the work of Frank Baginski, Chris Jenkins and Bob Walls.  Prototype
vessels were inflated at the Princeton Plasma Physics
Laboratory~\cite{pppl-news}, and later at the Princeton University Department
of Athletics' Jadwin Gym, with the kind permission of both organizations.  The
Honeywell (Pottsville, PA) and mf-folien (Germany) extrusion plants permitted
our inspections and implemented our suggested cleanliness precautions with more
willing cooperation on their parts than we had any right to expect.  We could
not have selected and purified our materials without the work of Tama
Chemicals, CleanFilm, and many other companies and institutions.

Funding for this paper and for the activities reported herein were provided by
the National Science Foundation (grants \mbox{PHY-0077423}, \mbox{PHY-0201141},
\mbox{PHY-0503816}), and by the Instituto Nazionale di Fisica Nucleare.
Co-author L.\ Cadonati gratefully acknowledges the support of NSF grant
\mbox{PHY-0107417}.  Co-author R.\ B.\ Vogelaar gratefully acknowledges the
support of NSF grants \mbox{PHY-9972127} and \mbox{PHY-0501118}.

\newpage

\newpage

\newpage

\newpage

\newpage

\newpage

\newpage

\newpage

\newpage

\newpage

\newpage

\newpage

\newpage

\newpage

\newpage

\end{document}